\documentclass[10pt]{article}
\usepackage[latin1]{inputenc}
\usepackage{amsmath}
\usepackage{amsfonts}
\usepackage{amssymb}
\usepackage{subfig}
\usepackage{fancybox,graphicx}
\usepackage{subfig}
\usepackage{color}
\usepackage{authblk}

\usepackage[top=2in, bottom=1.5in, left=1in, right=1in]{geometry}

\title{Exact Energy Computation of the One Component Plasma on a Sphere for Even Values of the Coupling Parameter} 

\author[*,$\dag$]{R. Salazar}
\author[*]{G. T\'ellez}
\affil[*]{Departamento de F\'isica, Universidad de los Andes - Bogot\'a, Colombia}
\affil[$\dag$]{Laboratoire de Physique Th\'eorique (UMR 8627), Universit\'e de Paris-Sud and CNRS, B\^atiment 210, 91405 Orsay Cedex, FRANCE}
\begin{document}
\maketitle

\begin{abstract}
The two dimensional one component plasma 2dOCP is a classical system
consisting of $N$ identical particles with the same charge $q$
confined in a two dimensional surface with a neutralizing
background. The Boltzmann factor at temperature $T$ may be expressed
as a Vandermonde determinant to the power $\Gamma=q^2/(k_B
T)$. Several statistical properties of the 2dOCP have been studied by
expanding the Boltzmann factor in the monomial basis for even values
of $\Gamma$. In this work, we use this formalism to compute the energy
of the 2dOCP on a sphere. Using the same approach the entropy is
computed. The entropy as well as the free energy in the thermodynamic
limit have a universal finite-size correction term $\frac{\chi}{12}\log N$, where $\chi=2$ is the Euler characteristic of the sphere. A non-recursive formula for coefficients of monomial functions expansion is used for exploring the energy as well as structural properties for sufficiently large values of $\Gamma$ to appreciate the crystallization features for $N=2,3,\ldots,9$ particles. Finally, we make a brief comparison between the exact and numerical energies obtained with the Metropolis method for even values of $\Gamma$.
\end{abstract}

\section{Introduction}
Coulomb systems, plasmas and electrolytes are systems of charged
particles interacting according to the Coulomb's law. The
One-Component Plasma (OCP) or jellium is the simplest model of a
Coulomb system. It is a set of $N$ identical pointlike particles of
charge $q$ embedded in a neutralizing background. The pair potential
between particles is the solution of the Poisson equation. This
solution is logarithmic for a system in two dimensions. The only
dimensionless coupling constant is $\Gamma = q^2/(k_B T)$ where $k_B$
is the Boltzmann constant and $T$ is the temperature. In general, a
two dimensional OCP (2dOCP) with logarithmic interaction does not
describe real (three-dimensional) charged particles confined on a surface because they interact with the usual inverse power law potential. However, this logarithmic case has been widely studied because it offers analytic solutions on diverse geometries particularly for $\Gamma=2$ \cite{sari,sari2,ginibre,jancoviciDisk}. At the special coupling $\Gamma = 2$ where the 2dOCP is in the fluid phase the distribution functions may be found exactly. In particular, the pair correlation function $g(r)$ is reduced to a gaussian form $\exp(-\pi\rho r^2)$ with $\rho$ the density. Expansions around $\Gamma=2$ suggests that the pair correlation function changes from the exponential form to an oscillating one for a region with $\Gamma>2$. This behaviour of the pair correlation function as the coupling is stronger has been observed in Monte Carlo simulations \cite{MontecarloStudyCaillol}\cite{ChoquardClerouin1983}. For sufficient high values of $\Gamma$ (low temperatures) the 2dOCP begins to crystallize and there are several works where the freezing transition is found. For the case of the sphere Caillol  et al. \cite{MontecarloStudyCaillol} localized the  coupling parameter for melting at $\Gamma \approx 140$.

In the limit $\Gamma\rightarrow\infty$ the 2dOCP becomes a Wigner crystal. In particular, the spatial configuration of the charges which minimizes the energy at zero temperature for the 2dOCP on a plane is the usual hexagonal lattice. Nowadays, the corresponding Wigner crystal of the 2dOCP on sphere or Thomson problem may be solved numerically \cite{Mughal2013} and the unsolved analytical problem has been included in the Smale's list of problems for the 21st century. Experimentally, the dusty plasmas are one of the ways to obtain a bidimensional Coulomb system on the plane. A plasma is an ionised gas with a low ionisation degree range frequently referred as the fourth state of matter. If the Coulomb interaction energy of the particles is much higher than the individual kinetic energy, then the particles may arrange themselves in a lattice forming a crystal. Different research groups studied these plasma crystals in the laboratory during the nineties \cite{ChuAndLin,ThomasEtAl1994} by confining particles of several micrometers on horizontal layers whose observation is made by illuminating such plane with HeNe laser light. There are also experimental investigations on two-dimensional spherical crystals which are formed in the surface of water droplets in oil where defects of the ideal crystal as disclinations and dislocations are observed in the laboratory \cite{Bausch2003}.

The main objective of this work is to make exact computations of the
energy and the entropy of the 2dOCP on a sphere at
$\Gamma \neq 2$ and reproduce the well known results at
$\Gamma=2$. Previously, a converging series expansions of the truncated pair correlation function on the disk at $\Gamma=4$ was described in \cite{samajGamma4}. For the case of the sphere it is possible to implement an extension of the
expansion Vandermonde determinant to the power $\Gamma$ techniques
used in the analytical computation of the free energy presented in
\cite{TellezForrester1999,TellezForrester2012}. The results of those
works give the free energy for fixed values of the temperature,
therefore it is not straightforward to obtain the internal energy from
those results. Nevertheless, in this work we will obtain the internal
energy by using the relation that expresses it in terms of the pair
correlation function. Once the internal energy is known, the entropy
would be found straightforwardly from the free energy and the internal
energy.

This document is organized as follows. In the next section a
description of the system and main definitions will be
done. Previously, the authors of \cite{TellezForrester2012} were able
to obtain the partition $Z_{N,\Gamma}$ and pair correlation function
$\rho^{(2)}(\theta)$ constrained to the condition $\Gamma=q^2/(k_B
T)=2,4,...,2n$ with $n$ a positive integer for several geometries
including the sphere. We will use some of these results to compute the
energy and entropy under the same restriction over coupling
parameter. Section \ref{sec:vandermonde} summarizes the basic technique behind the
computation of $Z_{N,\Gamma}$ and $\rho^{(2)}(\theta)$ and defines the
notation adopted until the end of the document. Sections
\ref{sec:energy} and \ref{sec:entropy} are devoted to describe the exact energy and
entropy computations including a comparison with simulation results
obtained with the Metropolis method. The analytic method described in
this document is mostly based in the expansion of the Vandermonde
determinant to the power $\Gamma$ using the monomial functions
$m_{\mu}(z_1,\ldots,z_N)$ as a basis where $z_1,\ldots,z_N$ are
related with the particle's positions and $\mu$ labels each monomial
function. Since the energy and entropy will be expressed as expansions
over the symmetric and antisymmetric monomial functions, then
two-different techniques for computing the expansions coefficients
$\{C_{\mu}^{(N)}(\Gamma/2)\}$ are included in the appendix section.
The first technique uses the multinomial expansion theorem to find
these coefficients. The second technique is based in the partial
derivatives differentiation of the Vandermonde determinant to the
power $\Gamma$ combined with the Finite Difference Method (FDM) in
order to obtain $C_{\mu}^{(N)}(\Gamma/2)$. Both methods give the exact
value of coefficients including the one based on FDM because it uses
the fact that $n$-order finite difference of a polynomial of order $n$
is exact.

\section{System description}
This work will be focused in the study of the 2dOCP on a sphere. It is a classical system of $N$ particles with charge $q$ on a two dimensional surface with a neutralizing background $\rho_b$. The Coulomb interaction potential $\nu(\vec{r}_1,\vec{r}_2)$ between two particle located at $\vec{r_1}$ and $\vec{r_2}$ with respect the center of the sphere is 

\[
\nu(\vec{r}_1,\vec{r}_2) = -\log\left(\frac{\left|\vec{r}_1-\vec{r}_2\right|}{L}\right)
\]\\
where $L$ is an arbitrary parameter which defines the length scale. In this writing we will study the 2OCP on a sphere. The \textit{excess energy} $U_{exc}$ of the 2OCP is given by

\[
U_{exc} = U_{pp} + U_{bp} + U_{bb}
\]\\ 
where $U_{pp}$ is the \textit{particle-particle interaction energy} contribution

\begin{equation}
U_{pp} = -q^2\sum_{1 \leq i<j\leq N}\log\left(\frac{\left|\vec{r}_i-\vec{r}_j\right|}{L}\right),
\label{UppSimpleDefintionEq}
\end{equation}
the \textit{background-particle interaction} contributes with an energy 
\[
U_{pp} = q\sum_{i=1}^N V_b\left(\vec{r}_i\right),
\]
\\where 
\[
V_b\left(\vec{r}\right)=\int_{sphere}\log\left(\frac{\left|\vec{r}-\vec{r}'\right|}{L}\right)\rho_b(\vec{r}')dS'
\]
\\is the background potential with $dS'$ the area element of the sphere. Finally, the background also interacts with itself and contributes with an energy 

\[
U_{bb} =-\frac{q^2}{2}\int_{sphere}\rho(\vec{r})V_b(\vec{r})dS\,.
\]
If the background density is constant $\rho_b=\frac{N}{4\pi R^2}$ with $R$ the radius of the sphere, then the particle-background and background-background interactions may be computed directly from their definitions. For the sphere they are

\[
U_{pb} = \frac{q^2 N^2}{2}\left[ 2\log\left(\frac{2R}{L}\right)-1\right]\hspace{1.0cm} \mbox{and} \hspace{1.0cm}U_{bb} = -\frac{q^2 N^2}{4}\left[ 2\log\left(\frac{2R}{L}\right)-1\right].
\]
The particle-particle interaction energy requires a detailed treatment that will be discussed in this writing. The $U_{pp}$ average contribution may be computed by using the following equation 

\begin{equation}
<U_{pp}> = \frac{1}{2}\int \rho^{(2)}(\vec{r}_1, \vec{r}_2) \nu(\vec{r}_1, \vec{r}_2)d\vec{r}_1 d\vec{r}_2
\label{UppEnergyDefinitionEq}
\end{equation}
where the integrations are over the sphere and $\rho^{(2)}(\vec{r}_1, \vec{r}_2)$ is the \textit{two-point correlation function}. Since the partition function may be computed for even values of the coupling parameter and $\rho^{(2)}(\vec{r}_1, \vec{r}_2)$ is found by functional derivatives of the partition function, then it is possible to compute Eq.~(\ref{UppEnergyDefinitionEq}) and the excess energy. 
\section{Vandermonde expansion approach for the 2dOCP on the sphere}
\label{sec:vandermonde}
\subsection{The partition function}
The canonical partition function $Z_c$ is 

\[
Z_c(T,A,N) :=  \left(\frac{2m\pi k_B T}{h^2}\right)^N Z_{N,\Gamma}
\]
\\where $A=4\pi R^2$ is the area of the sphere, $m$ the mass of the
particles, $h$ is Planck constant and $Z_{N,\Gamma}$ is the configurational partition function 
\[
Z_{N,\Gamma} := \frac{1}{N!} \int_{Sphere^N}dS_1\cdots dS_N \exp\left(-\beta U_{exc}\right)  \hspace{1.0cm}\mbox{with}\hspace{1.0cm}\int_{Sphere}dS_i = \int_0^{\pi}\int_0^{2\pi} R^2\sin(\theta)d\theta d\phi
\]
\\and 
\begin{equation}
U_{exc} = -\frac{\partial \log Z_{N,\Gamma}}{\partial\beta} = <U_{pp}> + \frac{q^2 N^2}{4}\left[ \log\left(\frac{N}{\rho_b \pi L^2}\right)-1\right].
\label{excesEnergyEq}
\end{equation}
\\The total energy is the usual bidimensional ideal gas  energy plus the excess energy contribution
\begin{equation}
U = -\frac{\partial \log Z_c(T,A,N)}{\partial\beta} = N k_B T + U_{exc}.
\label{totalEnergy}
\end{equation}
\\Now, the Coulomb potential may be written as follows \cite{Caillol} 

\[
\nu(\theta_1,\phi_1,\theta_2,\phi_2) = -\log\left(\frac{2R}{L}\left| u_2v_1-u_1v_2\right|\right),
\]
\\where $R$ is the radius of the sphere, $\theta$ and $\phi$ are the usual angles of spherical coordinates and  $u(\theta,\phi)=\cos\left(\frac{\theta}{2}\right)\exp\left(i\frac{\phi}{2}\right)$, $v(\theta,\phi)=-i\sin\left(\frac{\theta}{2}\right)\exp\left(i\frac{\phi}{2}\right)$ are the Cayley-Klein parameters. Hence the Boltzmann factor takes the form
\[
\exp\left(-\beta U_{exc}\right) = \left(\frac{L}{2R}\right)^{\frac{\Gamma N}{2}}\exp\left(\frac{\Gamma N^2}{4}\right)\prod_{1 \leq i<j\leq N} \left| u_jv_i-u_iv_j\right|^\Gamma.
\]
\begin{figure}[h]
  \centering   
  \includegraphics[width=0.6\textwidth]{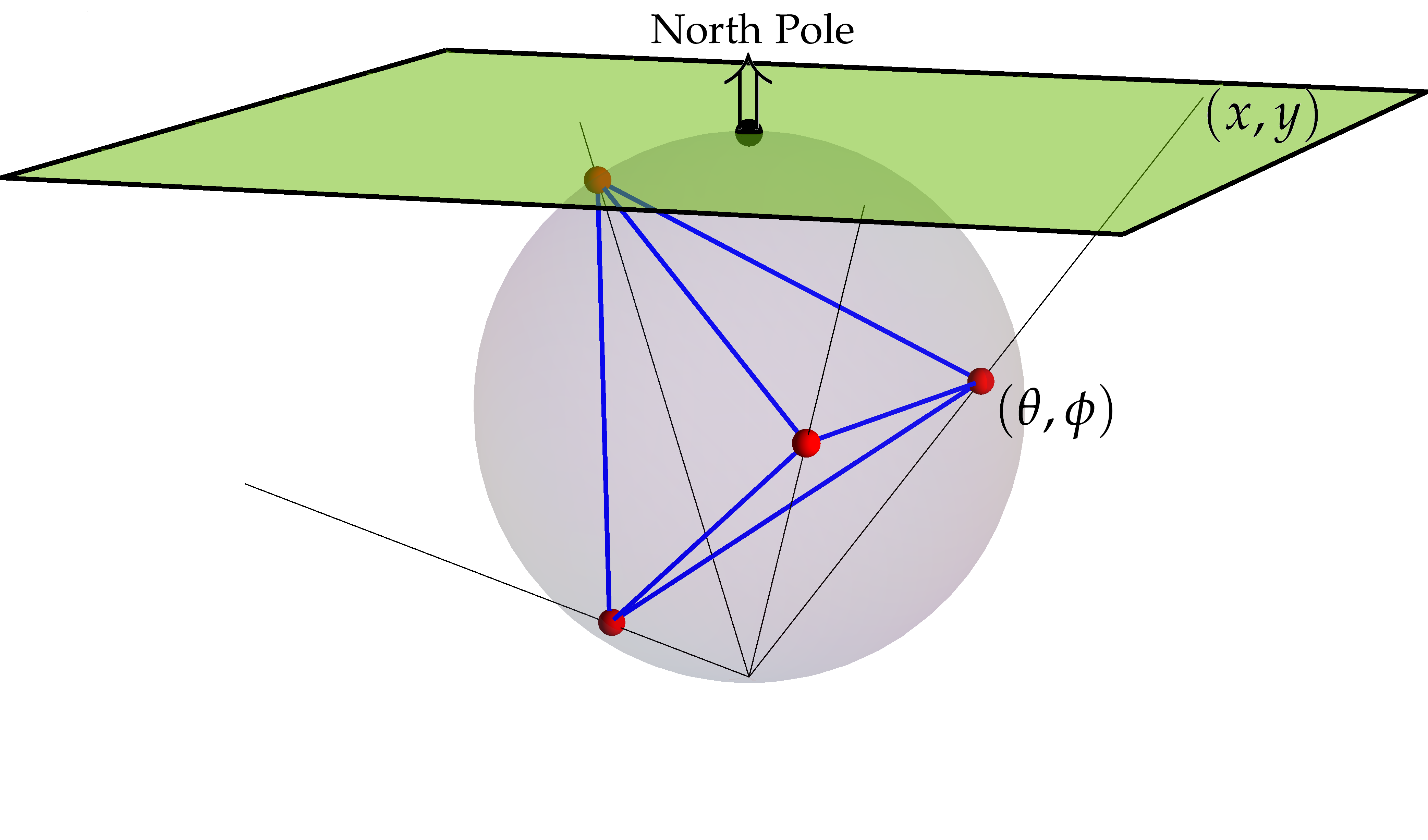}
  \caption[Stereographic projection.]%
  {Stereographic projection.}
\end{figure}
\\It is convenient to apply a stereographic projection from the angles $(\theta, \phi)$ to the coordinates  $(x, y)$ on the plane tangent to the north pole of the sphere. If the complex variable $z = x+iy$ is defined, then it may be written as $z = 2R\exp(i\phi)\tan\left(\frac{\theta}{2}\right)$ so the Boltzmann factor takes the form
\[
\exp\left(-\beta U_{exc}\right) = \left(\frac{L}{2R}\right)^{\frac{\Gamma N}{2}}\exp\left(\frac{\Gamma N^2}{4}\right)\prod_{i=1}^N \left(\frac{1}{1+\frac{|z_i|^2}{4R^2}}\right)^{\frac{\Gamma(N-1)}{2}} \prod_{1 \leq i<j\leq N} \left|\frac{z_i-z_j}{2R}\right|^\Gamma .
\]
\\On the other hand, the integrals over the surfaces in the new variables are 
\[
\int_{Sphere^N}dS_1\cdots dS_N \rightarrow  \int_{\Re^{2N}}\prod_{i=1}^N \frac{d\vec{r}_i}{(1+\frac{|z_i|^2}{4R^2})^2}
\]
\\where $d\vec{r}_i=dx_i dy_i$ is the area element on the projected plane. As a result, the partition function may be written as follows
\[
Z_{N,\Gamma} = \left(\frac{L}{2R}\right)^{\frac{\Gamma N}{2}}\exp\left(\frac{\Gamma N^2}{4}\right) \frac{1}{N!} \int_{(\Re^2)^N}  \prod_{1 \leq i<j\leq N} \left|\frac{z_i-z_j}{2R}\right|^\Gamma \prod_{i=1}^N \frac{d\vec{r}_i}{\left(1+\frac{|z_i|^2}{4R^2} \right)^{\frac{\Gamma(N-1)}{2}+2 } } .
\]
The difficulties in the integration especially rises from the product $\prod_{1 \leq i<j\leq N} \left|\frac{z_i-z_j}{2R}\right|^\Gamma$. For $\Gamma=2$ such product may be written in terms of the Vandermonde determinant $ \det(z_j^{i-1})_{(i,j=1,2,\ldots,N)}$ as follows $\prod_{1 \leq i<j\leq N} \left|\frac{z_i-z_j}{2R}\right|^2 =\frac{1}{(2R)^N} \det(z*_j^{i-1})\det(z_j^{i-1})$ and the partition function may be found explicitly. The exact solution of the problem for even values of $\Gamma$ was obtained by the authors of \cite{TellezForrester1999, TellezForrester2012} where $Z_{N,\Gamma}$ was found by using the following expansion

\begin{equation}
\prod_{1 \leq i<j\leq N} \left(z_i-z_j\right)^{\Gamma/2} = \sum_{\mu}C_{\mu}^{(N)}(\Gamma/2)m_\mu(z_1,\ldots,z_N).
\label{expansionEq}
\end{equation}
\\The indices set $\mu:=(\mu_1,\ldots,\mu_N)$ is a partition of
$\Gamma N(N-1)/4$ with the condition
$(N-1)\Gamma/2\geq\mu_1\geq\mu_2\cdots\geq\mu_N\geq 0$ for
\textbf{even values} of $\Gamma/2$ and a partition of $\Gamma
N(N-1)/4$ with the condition
$(N-1)\Gamma/2\geq\mu_1>\mu_2\cdots>\mu_N\geq 0$ for \textbf{odd
  values} of $\Gamma/2$. The $m_\mu(z_1,\ldots,z_N)$ are the monomial
symmetric or antisymmetric functions, depending on the parity of $\Gamma/2$, 
\[
m_{\mu}(z_1,\ldots ,z_N) = \frac{1}{\prod_i m_i !} \sum_{\sigma\in S_N} \mbox{sign}(\sigma)^{b(\Gamma)} z_{\sigma_1}^{\mu_1}\cdots z_{\sigma_N}^{\mu_N}
\]
where the sum is over the permutations of a given partition $\mu_1,\ldots,\mu_N$, the variable $m_i$ denotes the frequency of the index $i$ in such partition (one for the odd values of $\Gamma/2$) and $b(\Gamma)$ is defined as
\[
b(\Gamma) = \left\{
\begin{array}{rl}
1 & \mbox{ if } \Gamma/2 \mbox{ is odd } \\
0 & \mbox{ if } \Gamma/2 \mbox{ is even }
\end{array}
\right. .
\]
For integer values of $\Gamma/2$ the coefficient $C_{\mu}^{(N)}(\Gamma/2)$ takes an integer value depending on $N$ and $\mu$. If the expansion given by Eq.~(\ref{expansionEq}) is used, then the partition function takes the form
\[
Z_{N,\Gamma} = \left(\frac{L}{2R}\right)^{\frac{\Gamma N}{2}}\exp\left(\frac{\Gamma N^2}{4}\right) \left(4\pi R^2\right)^N\frac{1}{N!} \sum_{\mu}\frac{\left[C_{\mu}^{(N)}(\Gamma/2)\right]^2}{\prod_i m_i !}\prod_{l=1}^N G_{\mu_l}\left[ g(\tilde{r}^2) \right]
\]
\\where 
\[
G_{\mu_l}[g(r^2)] := 2\int_0^\infty dr\, r^{1+2\mu_l}g(r^2)  \hspace{1.0cm}\mbox{with}\hspace{1.0cm}  g(r^2) = \frac{1}{\left(1+r^2\right)^{2+\Gamma(N-1)/2}}
\]
and $\tilde{r}^2=|\frac{z}{2R}|^2=\tan\left(\frac{\theta}{2}\right)$. The integral function $G$ may be computed by using $\int_0^\infty du \frac{u^p}{(1+u)^q} = \frac{p!(q-p-2)!}{(q-1)!}$. Finally, the partition function is 

\begin{equation}
Z_{N,\Gamma} = \left(\frac{L}{2R}\right)^{\frac{\Gamma N}{2}}\exp\left(\frac{\Gamma N^2}{4}\right) \left[\frac{4\pi R^2}{\left(\frac{\Gamma}{2}(N-1)+1\right)!}\right]^N Z_{N,\Gamma}^{sphere}
\label{cofigurationalPartitionFunctionEq}
\end{equation}
where

\[
Z_{N,\Gamma}^{sphere} = \sum_{\mu}\frac{\left[C_{\mu}^{(N)}(\Gamma/2)\right]^2}{\prod_i m_i !}\prod_{l=1}^N \mu_l!\left((N-1)\Gamma/2-\mu_l\right)! .
\]
\subsection{Density and pair correlation function}
If the following partition function is defined as
\begin{equation}
Z_{N,\Gamma}[f] := \frac{1}{N!} \int_{Sphere^N}dS_1\cdots dS_N \exp\left[-\beta \left(U_{exc} + \sum_{i=1}^N f(\vec{r}_i)\right)\right]
\label{partionFunctionWithFieldEq}
\end{equation}
with $f(\vec{r})$ an arbitrary function, then it is well known that density $n^{(1)}(\vec{r})$ may be found by a functional derivation of the partition function \begin{equation}
-\beta n^{(1)}(\vec{r}) = \frac{1}{Z_{N,\Gamma}[f]}\frac{\delta Z_{N,\Gamma}[f]}{\delta f(\vec{r})} =\frac{1}{Z^{sphere}_{N,\Gamma}[f]} \frac{\delta Z^{sphere}_{N,\Gamma}[f]}{\delta f(\vec{r})}
\label{densityGeneralEq}
\end{equation}
where 
\[
Z_{N,\Gamma} = \left(\frac{L}{2R}\right)^{\frac{\Gamma N}{2}}\exp\left(\frac{\Gamma N^2}{4}\right) \left(2\pi R\right)^N\frac{1}{N!} \sum_{\mu}\frac{\left[C_{\mu}^{(N)}(\Gamma/2)\right]^2}{\prod_i m_i !}\prod_{l=1}^N G_{\mu_l}\left[ g(\tilde{r}^2)e^{\beta f(\vec{\tilde{r}})} \right].
\]
Similarly, the Ursell function $U^{(2)T}(\vec{r}_1,\vec{r}_2) = \rho^{(2)}(\vec{r}_1, \vec{r}_2) - n^{(1)}(\vec{r}_1)n^{(1)}(\vec{r}_2)$ may be found from 
\begin{equation}
-\beta^2 U^{(2)T}(\vec{r}_1,\vec{r}_2) = \frac{\delta^2 \ln Z_{N,\Gamma}[f]}{\delta f(\vec{r}_1) \delta f(\vec{r}_2)}
\label{ursellEq}
\end{equation}
where $\rho^{(2)}(\vec{r}_1, \vec{r}_2)$ is the pair correlation function. Using Eqs. (\ref{densityGeneralEq}) and (\ref{partionFunctionWithFieldEq}) the density takes the form
\begin{multline}
n^{(1)}(\theta) = \frac{\rho_b\left[(N-1)\Gamma/2+1\right]!}{N
  Z^{sphere}_{N,\Gamma} (1+\tilde{r}^2)^{(N-1)\Gamma/2} }
\sum_{\mu}\frac{\left[C_{\mu}^{(N)}(\Gamma/2)\right]^2}{\prod_i m_i !} \\
\times \prod_{l=1}^N \mu_l!\left((N-1)\Gamma/2-\mu_l\right)!\sum_{k=1}^N\frac{\tilde{r}^{2\mu_k}}{\left((N-1)\Gamma/2-\mu_k\right)!}. 
\label{probabilityDensityEq}
\end{multline}
We are free to put one of these charges on the north pole $\tilde{r}(0) = 0$ and $n^{(1)}(0)=\rho_b$ is a constant because of the sphere symmetry. Hence
\begin{equation}
1 = \left.\frac{\left[(N-1)\Gamma/2+1\right]!}{N Z^{sphere}_{N,\Gamma} (1+\tilde{r}^2)^{(N-1)\Gamma/2} } \sum_{\mu}\frac{\left[C_{\mu}^{(N)}(\Gamma/2)\right]^2}{\prod_i m_i !}\prod_{l=1}^N \mu_l!\left((N-1)\Gamma/2-\mu_l\right)!\sum_{k=1}^N\frac{\tilde{r}^{2\mu_k}}{\left((N-1)\Gamma/2-\mu_k\right)!}\right|_{ \tilde{r} \rightarrow 0}. 
\label{auxiliarEquation}
\end{equation}
Only the terms with $\mu_N=0$ contribute to the sum of
Eq.~(\ref{auxiliarEquation}) as a consequence of the limit
$\tilde{r}\to 0$. Using this condition we may write
\begin{equation}
Z_{N,\Gamma}^{sphere} =  \frac{\left[(N-1) \Gamma /2+1\right]!}{N}
\underset{ \tiny \hspace{0.1cm}\mbox{with}\hspace{0.1cm}
  \mu_N=0}{\sum_{\mu}}
\frac{\left[C_{\mu}^{(N)}(\Gamma/2)\right]^2}{\prod_i m_i
  !}\prod_{l=1}^{N-1} \mu_l!\left((N-1)\Gamma/2-\mu_l\right)!
\ .
\label{zSphereWithMuNZeroEq}
\end{equation}
Finally, the pair correlation function $\rho^{(2)}(\tilde{r})$ is computed by using Eqs. (\ref{ursellEq}), (\ref{probabilityDensityEq}) and (\ref{zSphereWithMuNZeroEq}). The result is the following  
\begin{multline}
\rho^{(2)}(\tilde{r}) =
\frac{\rho_b^2\left[(N-1)\Gamma/2+1\right]!^2}{N^2
  Z^{sphere}_{N,\Gamma} (1+\tilde{r}^2)^{(N-1)\Gamma/2} }
\underset{\tiny \hspace{0.1cm}\mbox{with}\hspace{0.1cm}
  \mu_N=0}{\sum_{\mu}}
\frac{\left[C_{\mu}^{(N)}(\Gamma/2)\right]^2}{\prod_i m_i !} \\
\times \prod_{l=1}^{N-1} \mu_l!\left((N-1)\Gamma/2-\mu_l\right)!\sum_{k=1}^{N-1}\frac{\tilde{r}^{2\mu_k}}{\mu_k!\left((N-1)\Gamma/2-\mu_k\right)!} 
\label{correlationFunctionEq} 
\end{multline}
where $\vec{r}_1$ was placed at the north pole of the sphere and $|\vec{r}_2|=\tilde{r}=\tan\left(\frac{\theta}{2}\right)$. 

\section{Energy}
\label{sec:energy}
The particle-particle interaction energy may be computed by using Eq.~(\ref{UppEnergyDefinitionEq}) with the pair correlation given by Eq.~(\ref{correlationFunctionEq}). It is suitable to set an additional polar system on the plane generated by the stereographic projection in order to evaluate Eq.~(\ref{UppEnergyDefinitionEq}), this is
\[
<U_{pp}> = \frac{1}{2} (4\pi R^2) \int_0^{2 \pi} d\phi_p \int_0^{\infty} \frac{r_p dr_p}{\left[1+\left(\frac{r_p}{2R}\right)^2\right]^2}\rho(r_p)\nu(r_p)
\]
where $z=x+iy$ with the projected variables $x=r_p\cos(\phi_p)$ and $y=r_p\sin(\phi_p)$. Here $r_p$ and $\phi_p$ play the role of polar coordinates in the plane of the stereographic projection. Defining $\tilde{r}=\frac{r_p}{2R} = \tan\left(\frac{\theta}{2}\right)$, then the Coulomb interaction potential between particles may be written as 

\[
\nu(\tilde{r}) = -\frac{1}{2}\log\left(\frac{\tilde{r}^2}{1+\tilde{r}^2}\right) - \log\left(\frac{2R}{L}\right)
\]
which splits the interaction energy in two parts
\[
<U_{pp}> = (4\pi R)^2 \left[\mathcal{I}_1 - \log\left(\frac{2R}{L}\right)\mathcal{I}_2\right]
\]
reducing the problem to compute these integrals
\[
\mathcal{I}_1 := \int_0^\infty\frac{\tilde{r}d\tilde{r}}{1+\tilde{r}^2}\rho^{(2)}(\tilde{r})\left[-\frac{1}{2}\log\left(\frac{\tilde{r}}{1+\tilde{r}^2}\right)\right] \hspace{1.0cm}\mbox{and}\hspace{1.0cm} \mathcal{I}_2 := \int_0^\infty\frac{\tilde{r}d\tilde{r}}{1+\tilde{r}^2}\rho^{(2)}(\tilde{r}).
\]
Using Eq.~(\ref{correlationFunctionEq}) the integral $\mathcal{I}_2$ takes the form 
\begin{multline}
  \mathcal{I}_2 = \frac{\rho_b^2\left[(N-1)\Gamma/2+1\right]!^2}{N^2
    Z^{sphere}_{N,\Gamma}  }
  \sum_{\mu}\frac{\left[C_{\mu}^{(N)}(\Gamma/2)\right]^2}{\prod_i m_i
    !}\prod_{l=1}^{N-1} \mu_l!\left((N-1)\Gamma/2-\mu_l\right)! \\
\times \sum_{k=1}^{N-1}\frac{1}{\mu_k!\left((N-1)\Gamma/2-\mu_k\right)!}\int_{0}^\infty\frac{\tilde{r}^{2\mu_k+1}}{ (1+\tilde{r}^2)^{(N-1)\Gamma/2}}. 
\end{multline}
This expression may be computed by using
\[
i^m_n := \int_{0}^\infty\frac{\tilde{r}^{2m+1}}{ (1+\tilde{r}^2)^n}d\tilde{r} = \frac{m!(n-m-2)!}{2 (n-1)!} 
\]
and the density relationship Eq.~(\ref{auxiliarEquation}). The result is
\[
 \mathcal{I}_2 = \frac{N-1}{2N}\rho_b^2 .
\]
On the other hand, the integral $\mathcal{I}_1$ takes the form
\begin{multline}
\mathcal{I}_1 = \frac{\rho_b^2\left[(N-1)\Gamma/2+1\right]!^2}{N^2 Z^{sphere}_{N,\Gamma}  } \sum_{\mu}\frac{\left[C_{\mu}^{(N)}(\Gamma/2)\right]^2}{\prod_i m_i !}\prod_{l=1}^{N-1} \mu_l!\left((N-1)\Gamma/2-\mu_l\right)! \\ \sum_{k=1}^{N-1}\frac{1}{\mu_k!\left((N-1)\Gamma/2-\mu_k\right)!}\int_{0}^\infty\frac{\tilde{r}^{2\mu_k+1}}{ (1+\tilde{r}^2)^{(N-1)\Gamma/2}}\left[-\frac{1}{2}\log(\frac{\tilde{r}}{1+\tilde{r}^2})\right] . 
\end{multline}
It includes the integral
\[
j^m_n := \int_{0}^\infty\frac{\tilde{r}^{2m+1}}{ (1+\tilde{r}^2)^n}\log(\frac{\tilde{r}}{1+\tilde{r}^2})d\tilde{r} 
\]
which may be found by using the following relationship
\[
\frac{\partial i^m_n}{\partial n} =  j^m_n +  i^m_n (H_{n-m-2}-H_{m}) 
\]
where $H_{m} = \sum_{k=1}^m \frac{1}{k}$ are the \textit{harmonic numbers}. Additionally, the integral $i^m_n$ is related with the \textit{beta function} 
\[B(x,y) = \int_0^1 t^{x-1}(1-t)^{y-1}dt \mbox{ for } \mbox{Re}(x),\mbox{Re}(y)>0\] 
as follows $i^m_n = \frac{1}{2}B(m+1,n-m-1)$. The derivative $\frac{d i^m_n}{dn}$ may be found by using the property $\frac{\partial B(x,y)}{\partial x} = B(x,y) \left[\psi(x)-\psi(x+y)\right]$ where 
\[
\psi(x) = \int_0^\infty \left( \frac{e^{-t}}{t} - \frac{e^{-x
    t}}{1-e^{-t}}  \right)\,dt
 \mbox{ for } \mbox{Re}(x)>0 
\]
is the \textit{digamma function}. If the argument is an integer, then $\psi(n)=-\gamma + H_{n-1}$ with $\gamma=0.577215664\ldots$ the \textit{Euler-Mascheroni constant}. So the partial derivative is $\frac{\partial i^m_n}{\partial n} = i^m_n (H_{n-m-2}-H_{n-1})$, hence  
\[
j^m_n = i^m_n\left(H_m-H_{n-1}\right) = \frac{m!(n-m-2)!}{2 (n-1)!}\left(H_m-H_{n-1}\right).
\]
Therefore, the integral $\mathcal{I}_1$ may be expressed as follows
\begin{multline}
\mathcal{I}_1 = \frac{\rho_b^2\left[(N-1)\Gamma/2+1\right]!}{4 N^2 Z^{sphere}_{N,\Gamma}  } \sum_{\mu}\frac{\left[C_{\mu}^{(N)}(\Gamma/2)\right]^2}{\prod_i m_i !}
\\
\times \prod_{l=1}^{N-1} \mu_l!\left((N-1)\Gamma/2-\mu_l\right)!\left\{ (N-1)H_{(N-1)\frac{\Gamma}{2}+1} - \sum_{m-1}^{N-1} H_{\mu_m} \right\}.
\end{multline}
Finally, the particle-particle interaction energy average is 
\begin{equation}
<U_{pp}> = \frac{Nq^2}{4}\left[(N-1)H_{(N-1)\frac{\Gamma}{2}+1} - \langle\sum_{m=1}^{N-1} H_{\mu_m}\rangle_{N-1}-(N-1)\log\left(\frac{N}{\rho_b\pi L^2}\right)\right]
\label{exactUppEq}
\end{equation}
where 
\begin{equation}
\langle\sum_{m=1}^{N-1} H_{\mu_m}\rangle_{N-1} = \frac{\sum_{\mu \hspace{0.1cm} with  \hspace{0.1cm} \mu_N=0 }\frac{\left[C_{\mu}^{(N)}(\Gamma/2)\right]^2}{\prod_i m_i !}\prod_{l=1}^{N-1} \mu_l!\left((N-1)\Gamma/2-\mu_l\right)!\left(\sum_{m=1}^{N-1} H_{\mu_m}\right) }{\sum_{\mu \hspace{0.1cm} with  \hspace{0.1cm} \mu_N=0 }\frac{\left[C_{\mu}^{(N)}(\Gamma/2)\right]^2}{\prod_i m_i !}\prod_{l=1}^{N-1} \mu_l!\left((N-1)\Gamma/2-\mu_l\right)!}\,,
\label{averageSumHEq}
\end{equation}
and the excess internal energy is
\begin{equation}
  U_{exc}=\frac{N q^2}{4} \left[ (N-1)H_{(N-1)\frac{\Gamma}{2}+1} -
    \langle\sum_{m=1}^{N-1}
    H_{\mu_m}\rangle_{N-1}-N+\log\left(\frac{N}{\rho_b\pi
      L^2}\right)\right]
\,.
\label{exactUexcAnyGammaEq}
\end{equation}

\subsection{Energy in the thermodynamic limit for $\Gamma=2$}
There is only one partition for $\Gamma=2$ associated to one coefficient $C_{\mu}^{(N)}(1)=1$. We are looking for partitions such that $\mu_N = 0$, then $\mu_i = N-i$ and $\prod_{i=1} m_i! = 1$ because the root partition does not repeat elements. As a result, the partition function is

\begin{equation}
Z^{(sphere)}_{2,N} = \prod_{i=1}^N (N-i)!(i-1)! = \left[\prod_{i=1}^N(i-1)!\right]^2.   
\label{zSphereGamma2Eq}
\end{equation}
\begin{figure}[h]
  \centering   
  \includegraphics[width=0.47\textwidth]{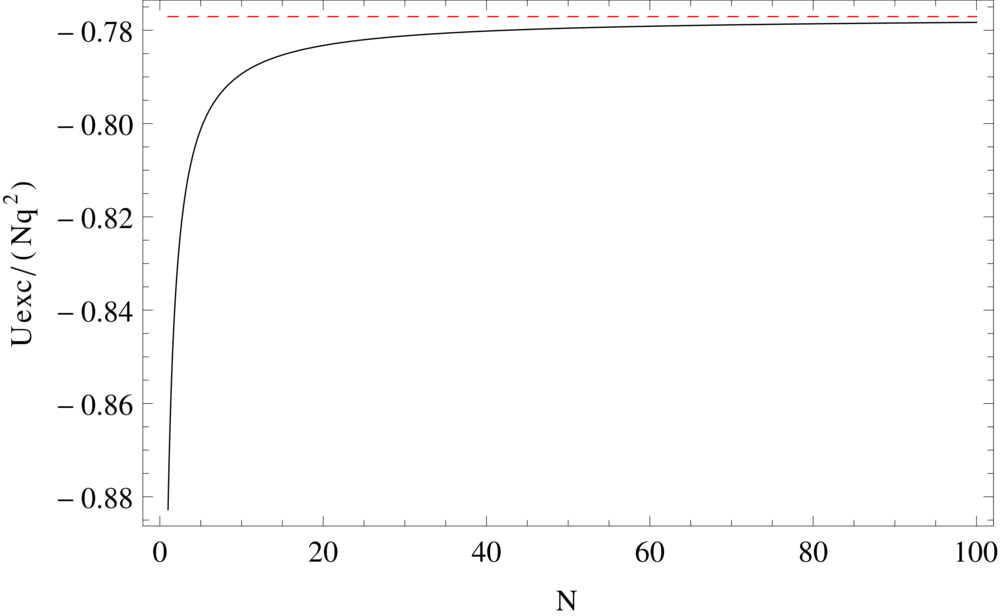}\hspace{0.2cm}
  \includegraphics[width=0.47\textwidth]{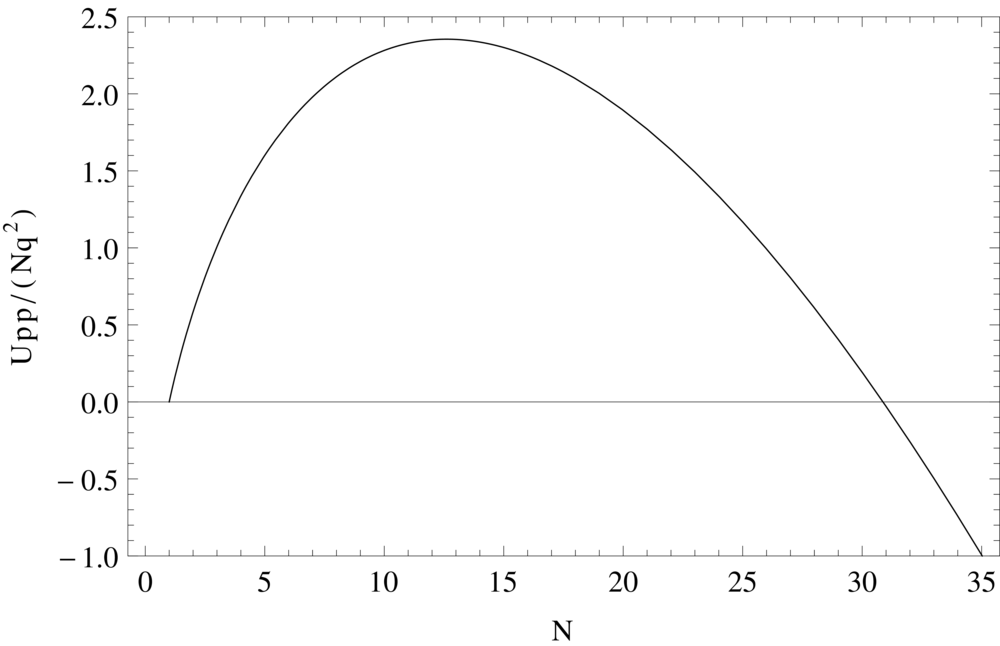}
  \caption[Energy for $\Gamma=2$.]%
  {Energy for $\Gamma=2$. \textbf{(left)} The solid line corresponds to the excess $U_{pp}$ per particle and squared charge and the solid line represents its value in the thermodynamic limit. \textbf{(right)} particle-particle interaction. In this plot we have set $\rho_b=1$ and $L=1$.}
\end{figure}
\\On the other hand, the average $\langle\sum_{m=1}^{N-1} H_{\mu_m}\rangle_{N-1}$ may be expressed as follows

\[
\langle\sum_{m=1}^{N-1} H_{\mu_m}\rangle_{N-1} = \sum_{m=1}^{N-1} H_{N-m} = NH_{N-1}-(N-1)
\]
where we have used the property $\sum_{u=1}^n H_{u} = (n+1)H_n-n$. Therefore, the particle-particle interaction energy average is
\[
<U_{pp}> = \frac{Nq^2}{4}\left[N-H_N-(N-1)\log\left(\frac{N}{\rho_b\pi L^2}\right)\right]
\]
and the excess energy given by Eq.~(\ref{excesEnergyEq}) is 
\begin{equation}
U_{exc} = \frac{Nq^2}{4}\left[\log\left(\frac{N}{\rho_b\pi L^2}\right)-H_N\right].
\label{UexcForGamma2Eq}
\end{equation}
This is just the result of \cite{MontecarloStudyCaillol}. The thermodynamic limit is defined by $N\rightarrow\infty$ while the area of sphere grows proportional to the number of particles in order to hold the density $\rho_b=N/(4\pi R^2)$ as a constant. The harmonic number for large $N$ is $H_N \underset{N\gg1}{\simeq} \gamma + \log N$ and the excess energy per particle is constant in this limit 
\begin{equation}
\lim_{N\rightarrow\infty} \frac{U_{exc}}{N} = -\frac{q^2}{4}\left[\log(\rho_b\pi L^2)+\gamma\right].
\label{UppPerParticleThermodinamicLimGamma2Eq}
\end{equation}
It coincides with the result for the 2DOCP in the bulk in a plane obtained by
B. Jancovici \cite{jancoviciDisk} who used a similar approach based on
the expansion of the free energy around $\Gamma - 2$. 

\subsection{Energy and pair correlation function for $N=2$}
\begin{figure*}
 \begin{minipage}[t]{.55\linewidth}\vspace{0pt}
  \includegraphics[width=\linewidth]{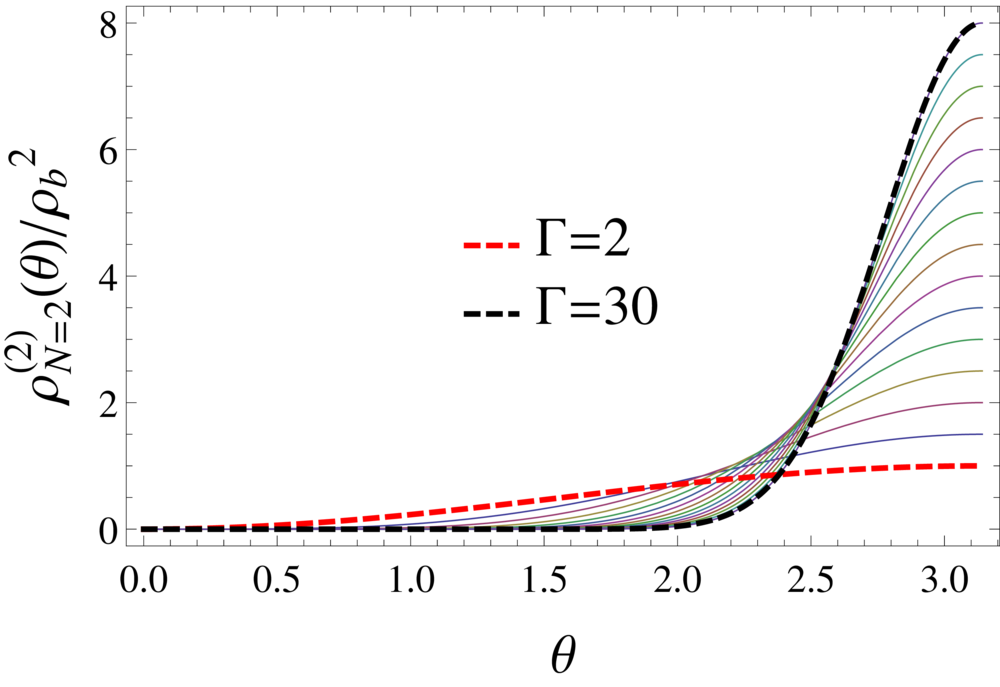}
  
 \end{minipage}
 \hspace{0.1cm}
 \hfill
 \begin{minipage}[t]{.3\linewidth}\vspace{0pt}\raggedright
  \begin{minipage}[t]{0.45\linewidth}\vspace{0pt}\raggedright
   \includegraphics[width=\linewidth]{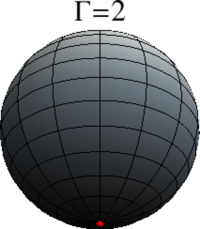}
   
  \end{minipage}\hfill
  \begin{minipage}[t]{0.45\linewidth}\vspace{0pt}\raggedright
   \includegraphics[width=\linewidth]{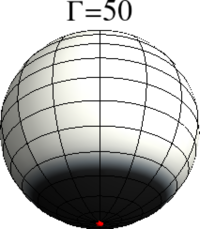}
   
  \end{minipage}\hfill
  \begin{minipage}[t]{0.45\linewidth}\vspace{0pt}\raggedright
   \includegraphics[width=\linewidth]{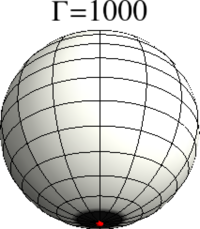}
   
  \end{minipage}\hfill
  \begin{minipage}[t]{0.45\linewidth}\vspace{0pt}\raggedright
   \includegraphics[width=\linewidth]{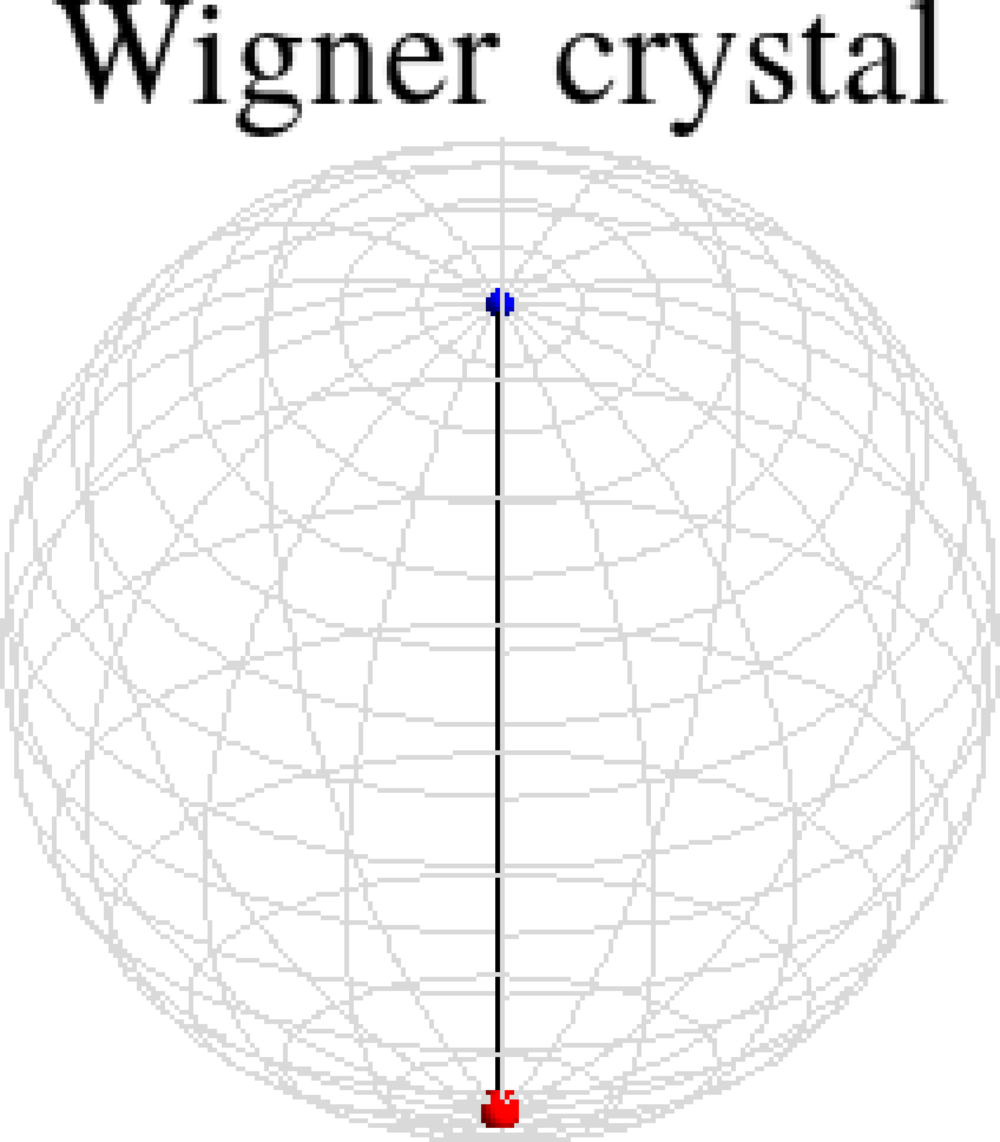}
   
  \end{minipage}
 \end{minipage}
\caption[Exact correlation function for $N=2$.]%
  {Exact correlation function of the 2dOCP for $N=2$ given by Eq(\ref{exactPairCorrelationForN2Eq}). \textbf{(left)} $\rho_{N=2}^{(2)}$ for several values of coupling parameter. The red and black dot-dashed lines corresponds to $\Gamma=2$ and $\Gamma=30$ respectively. Intermediate values $\Gamma=4,8,\ldots,28$ are represented by solid lines. \textbf{(right)} Density plots of $\rho_{N=2}^{(2)}$ and Wigner crystal for two particles. White color corresponds to $\rho_{N=2}^{(2)}(\theta)=0$. }  
\label{correlationFunctionForN2Fig}
\end{figure*}

If we consider the simple case of a pair of particles $N=2$, then the partitions are of the form $\mu=(\mu_1,\mu_2)$ where 

\[
\mu_1 = p, p+1, \cdots, 2p  \hspace{0.2cm}\mbox{ and }\hspace{0.2cm}  \mu_2 = p, p-1, \cdots, 0
\]
\\and $p= \Gamma / 4$. The coefficients are 
\[
C_\mu^{(2)}(2p) = (-1)^{\mu_1} \binom{2p}{\mu_1}.
\]
We also have included the derivation as a particular case for $N=2$ of the method described in section 7.2 (see Eq.~(\ref{coefficientsForN2ResultEq})) which is equivalent to apply the binomial theorem and the orthogonality condition of the monomial functions on Eq.~(\ref{expansionEq}). In fact, the idea to use the binomial theorem to find $C_\mu^{(2)}(\Gamma/2)$ is not new. The method described in section 7 is just a generalization of this idea with the aim to find $C_\mu^{(N)}(\Gamma/2)$ for other values of $N$ and $\Gamma$. Using these results for partitions and coefficients, then the partition function given by Eq.~(\ref{zSphereWithMuNZeroEq}) takes the form
\[
Z_{N,\Gamma}^{sphere} = \frac{\left[\Gamma /2+1\right]!}{2} \underset{\tiny \hspace{0.1cm}\mbox{with}\hspace{0.1cm} \mu_2=0}{\sum_{\mu}} \frac{1}{m_1!m_2!} \binom{\Gamma/2}{\mu_1}^2 \mu_1!\left((N-1)\Gamma/2-\mu_1\right)!.
\]
In general there are $p+1$ partitions, but only the partition $\mu=\left(\mu_1=2p,\mu_2=0\right)$ satisfy the condition $\mu_N=\mu_2=0$. The partition function for $N=2$ is 

\[
Z_{N,\Gamma}^{sphere} = \frac{1}{2}\left[\Gamma /2+1\right]!(\Gamma/2)!.
\]
Similarly, the harmonic numbers average defined in Eq.~(\ref{averageSumHEq}) is simply

\[
\langle\sum_{m=1}^{N-1} H_{\mu_m}\rangle_{N-1} = H_{\Gamma/2}
\]
\\so the particle-particle interaction energy average is 

\[
\left. < U_{pp} >\right|_{N=2} = \frac{q^2}{2}\left[\frac{1}{1+\Gamma/2} + \log\left(\frac{\rho_b\pi L^2}{2}\right)\right] 
\]
\\and the excess energy is 

\begin{equation}
\left. U_{exc}\right|_{N=2} = \frac{q^2}{2}\left[\frac{1}{1+\Gamma/2} - \log\left(\frac{\rho_b\pi L^2}{2}\right)-2\right]. 
\label{UexEnergyForN2eq}
\end{equation}
Curiously, this result coincides with the one found by derivation of the configurational partition function $U_{exc}=-\frac{\partial \log Z_{N,\Gamma}}{\partial \beta}=-q^2 -\frac{\partial \log Z_{N,\Gamma}}{\partial \Gamma}$ where  

\[
Z_{N=2,\Gamma} = (\pi L^2)^{\Gamma/2}\left(\frac{2}{\rho_b}\right)^{2-\Gamma/2}\frac{\exp(\Gamma)}{\Gamma+2},
\]
even when this function should be valid only for discrete even values of $\Gamma$ because it was found by assuming this condition. The pair correlation function
\begin{equation}
\rho_{N=2}^{(2)}(\theta) = \rho^2_b \frac{\Gamma/2+1}{2}\left(\frac{1-\cos\theta}{2}\right)^{\Gamma/2}
\label{exactPairCorrelationForN2Eq}
\end{equation}
was obtained straightforwardly from Eq.~(\ref{correlationFunctionEq}). Since the Wigner crystal for $N=2$ corresponds to the well known antipodal nodes configuration, then it is expected a concentration of the pair correlation function around $\theta=\pi$ as $\Gamma$ goes to infinity as it is shown in Fig. \ref{correlationFunctionForN2Fig}.   

\subsection{Energy and pair correlation function for $N=3$}
\begin{figure*}
 \begin{minipage}[t]{.55\linewidth}\vspace{0pt}
  \includegraphics[width=\linewidth]{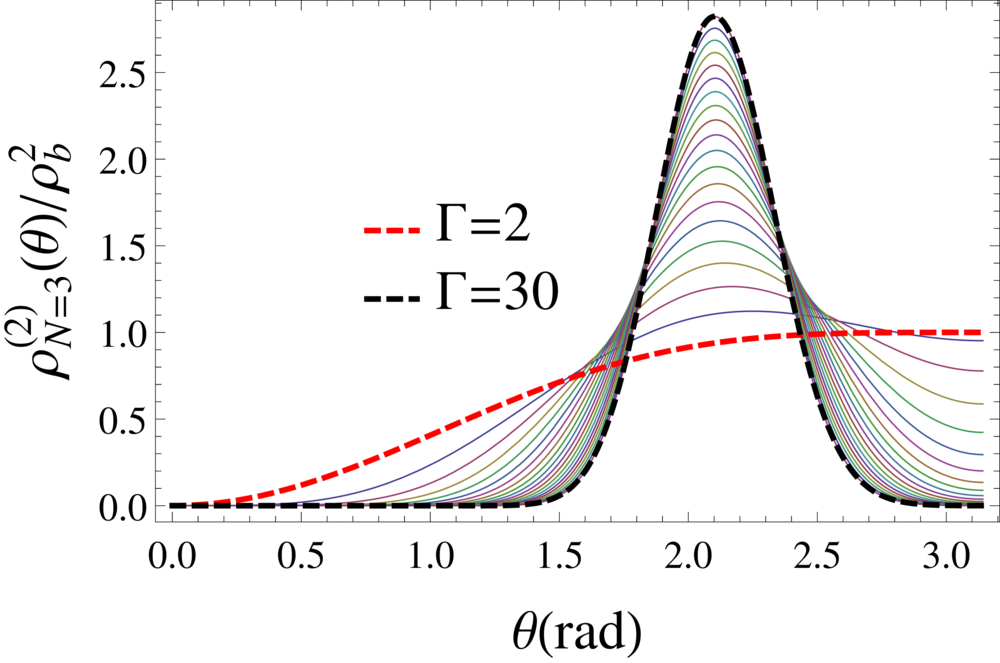}
  
 \end{minipage}
 \hspace{0.01cm}
 \hfill
 \begin{minipage}[t]{.4\linewidth}\vspace{0pt}\raggedright
  \begin{minipage}[t]{0.3\linewidth}\vspace{0pt}\raggedright
   \includegraphics[width=\linewidth]{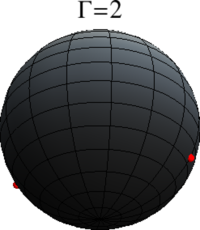}
   
  \end{minipage}\hfill
  \begin{minipage}[t]{0.3\linewidth}\vspace{0pt}\raggedright
   \includegraphics[width=\linewidth]{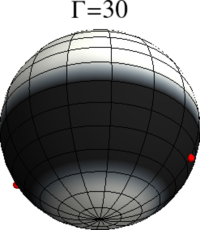}
   
  \end{minipage}\hfill
  \begin{minipage}[t]{0.3\linewidth}\vspace{0pt}\raggedright
   \includegraphics[width=\linewidth]{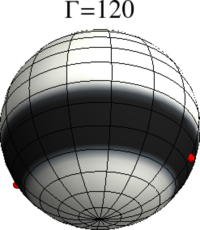}
   
  \end{minipage}\hfill
  \begin{minipage}[t]{0.3\linewidth}\vspace{0pt}\raggedright
   \includegraphics[width=\linewidth]{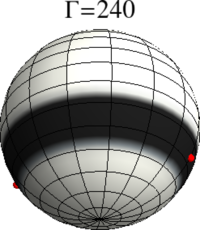}
   
  \end{minipage}\hfill
  \begin{minipage}[t]{0.3\linewidth}\vspace{0pt}\raggedright
   \includegraphics[width=\linewidth]{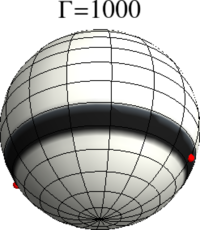}
   
  \end{minipage}\hfill
  \begin{minipage}[t]{0.3\linewidth}\vspace{0pt}\raggedright
   \includegraphics[width=\linewidth]{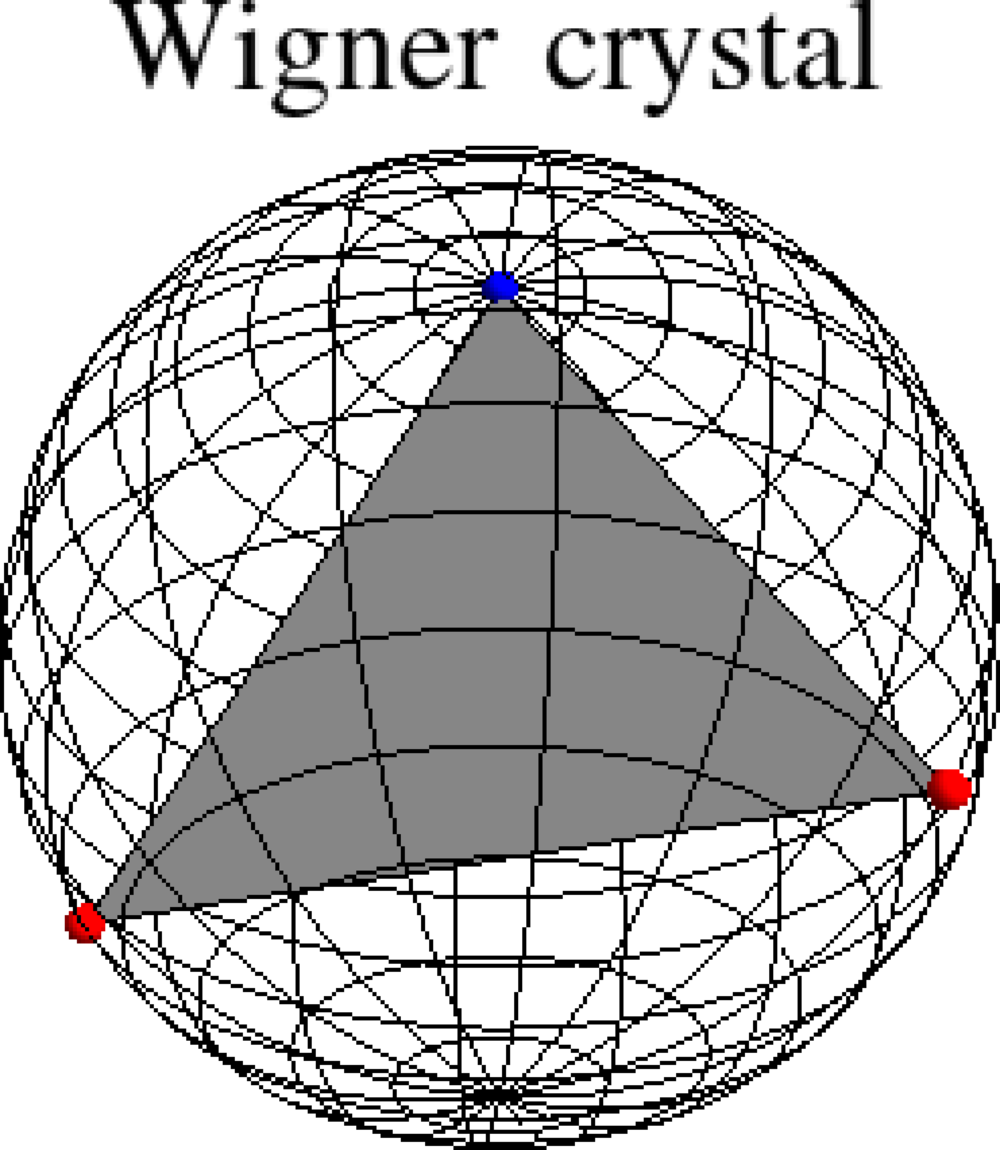}
   
  \end{minipage}
 \end{minipage}
\caption[Exact correlation function for $N=3$.]%
  {Exact correlation function of the 2dOCP for $N=3$. \textbf{(left)} $\rho_{N=3}^{(2)}$ for several values of coupling parameter. The red and black dot-dashed lines corresponds to $\Gamma=2$ and $\Gamma=30$ respectively. Intermediate values $\Gamma=4,8,\ldots,28$ are represented by solid lines. \textbf{(right)} Density plots of $\rho_{N=3}^{(2)}$ and Wigner crystal for three particles.}  
\label{correlationFunctionForN3Fig}
\end{figure*}

The excess energy as well as the pair correlation function may be obtained explicitly because the coefficients for three particles with $\mu_3=0$ are related with the well known coefficients for two particles without any restriction on partitions. The partitions and coefficients are given by
\[
\mu_1^{\alpha} = \Gamma + 1 - \alpha \mbox{,}\hspace{0.5cm} \mu_2^{\alpha} = \Gamma/2 + \alpha - 1 \hspace{0.5cm}\mbox{and}\hspace{0.5cm} \mu_3^{\alpha} = 0
\]
and
\[
C_\mu^{(3)}(\Gamma/2) =  (-1)^{\Gamma + 1 - \alpha} \binom{\Gamma/2}{\alpha-1} \hspace{0.5cm}\mbox{with}\hspace{0.5cm} \mu_3^{\alpha} = 0
\]
respectively. The derivation is done in appendix \ref{AppendixCoeffN3}. The index $\alpha$ is  introduced in order to count partitions running from 1 to $\mbox{Int}\left(\frac{\Gamma}{4}+1\right)$ the total number of partitions. Hence the pair correlation function of three particles takes the form
\begin{eqnarray}
\rho_{N=3}^{(2)}(\theta) = \frac{\rho^2_b (\Gamma+1)!}{3(1+\tan^2\left(\theta/2\right))^\Gamma}\frac{1}{\tilde{Z}_{3,\Gamma}}\sum_{\alpha=1}^{\mbox{Int}(\Gamma/4+1)}\frac{\binom{\Gamma/2}{\alpha-1}^2}{\left(\Pi_i m_i !\right)_{\alpha}}&&\left\{(\Gamma/2+\alpha-1)!(\Gamma/2-\alpha+1)!\left[\tan\left(\theta/2\right)\right]^{2(\Gamma+1-\alpha)} + \right. \nonumber \\ && \left.(\Gamma+1-\alpha)!(\alpha-1)!\left[\tan\left(\theta/2\right)\right]^{2(\Gamma/2+\alpha-1)}\right\}
\end{eqnarray}
where
\[
\tilde{Z}_{3,\Gamma} = \sum_{\alpha=1}^{\mbox{Int}\left(\frac{\Gamma}{4}+1\right)}\frac{1}{\left(\Pi_i m_i !\right)_{\alpha}} \binom{\Gamma/2}{\alpha-1}^2 \left[\left(\frac{\Gamma}{2}+\alpha-1\right)!\left(\frac{\Gamma}{2}-\alpha+1\right)!\tilde{r}^{2(\Gamma+1-\alpha)} + (\Gamma+1-\alpha)!(\alpha-1)!\tilde{r}^{2\left(\frac{\Gamma}{2}+\alpha-1\right)}\right]
\]
and 
\[\left(\Pi_i m_i !\right)_{\alpha} = \left\{ \begin{array}{rl}
 2 &\mbox{ if } \Gamma/2 \mbox{ is even and } \alpha= \mbox{Int}\left(\frac{\Gamma}{4}+1\right) \\
 1 &\mbox{ if } \Gamma/2 \mbox{ is odd }
       \end{array} \right.\]
is the multiplicity. A plot of the pair correlation function of three particles is shown in Fig. \ref{correlationFunctionForN3Fig}. For values of $\Gamma$ near to 2 the function $\rho_{N=3}^{(2)}$ is practically delocalized because of high thermal excitations. As the temperature is decreased to zero $\Gamma\rightarrow\infty$ the pair correlation function is concentrated around $\phi=2\pi/3$ radians obeying to the fact that particles crystallize in a equilateral triangle and the azimuthal symmetry of the system.

The excess energy may be obtained by using Eqs.~(\ref{averageSumHEq}), (\ref{exactUppEq}), and (\ref{excesEnergyEq}) with the explicit values of partitions and coefficients for $N=3$. The result is 
\[
\left.U_{exc}\right|_{N=3} = \frac{3}{4}q^2\left\{2H_{\Gamma+1}-\langle\sum_{m=1}^{2} H_{\mu_m}\rangle_{2}+\log\left(\frac{3}{\rho\pi L^2}\right)-3\right\}
\]
with
\[
\langle\sum_{m=1}^{2} H_{\mu_m}\rangle_{2} = \frac{1}{\tilde{Z}_{3,\Gamma}}\sum_{\alpha=1}^{\mbox{Int}\left(\frac{\Gamma}{4}+1\right)}\left(\frac{1}{\Pi_i m_i !}\right)_{\alpha} \binom{\Gamma/2}{\alpha-1}^2 \left[\left(\frac{\Gamma}{2}+\alpha-1\right)!\left(\frac{\Gamma}{2}-\alpha+1\right)!(\Gamma+1-\alpha)!(\alpha-1)!\right].
\]
\subsection{Asymptotic energy for $N \leq 4$}
\begin{figure}[h]
  \centering   
  \includegraphics[width=0.5\textwidth]{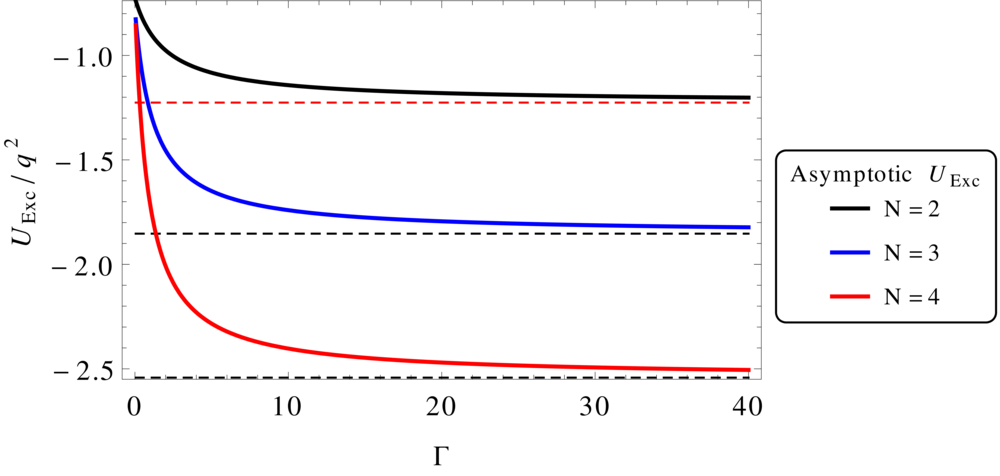}
  \caption[Asymptotic excess energy for $N \leq 4$.]%
  {Asymptotic excess energy for $N \leq 4$. The solid and dashed lines are the asymptotic behaviour of the excess energy given by Eq.~(\ref{UexcInTheAsymptoticLimitNleq4Eq}) and the Wigner crystal energy respectively. }
  \label{UexcInTheAsymptoticLimitNleq4Fig}
\end{figure}

The Wigner crystals for $N \leq 4$ at vanishing temperature play an special role because they are the only equidistant configurations of the Thomson problem. Since the cases $N=2$, $N=3$ and $N=4$ correspond to the antipodal nodes, equilateral triangle and the tetrahedron configurations  respectively, then we may compute the energy of these configurations by replacing the particle positions in Eq.~(\ref{UppSimpleDefintionEq}) and using Eq.~(\ref{excesEnergyEq}). The excess energy at $\Gamma\rightarrow\infty$ is

\[
\lim_{\Gamma \to \infty}\left.U_{exc}\right|_{N=2,3,4} = -(N-1)\frac{N q^2}{4}\left[\log\left(\frac{\eta_N^2}{4}\right)+\log\left(\frac{N}{\rho\pi L^2}\right)\right] + \frac{q^2 N^2}{4}\left[ \log\left(\frac{N}{\rho_b \pi L^2}\right)-1\right]
\]
\begin{equation}
\lim_{\Gamma \to \infty}\left.U_{exc}\right|_{N=2,3,4} = \frac{N q^2}{4}\left[(N-1)\log\left(\frac{4}{\eta_N^2}\right)+\log\left(\frac{N}{\rho\pi L^2}\right)-N\right] 
\label{UexcWignerNleq4Eq}
\end{equation}
\\where  $\eta_N$ is a geometrical factor which takes the values $\eta_2 = 2$, $\eta_3 = \sqrt{3}$ and $\eta_4 = \sqrt{\frac{8}{3}}$. Generally, the exact excess energy value of the Wigner crystal requires the evaluation of Eq.~(\ref{averageSumHEq}) at zero temperature. Unfortunately, the sum included in the average $\langle\sum_{m=1}^{N-1} H_{\mu_m}\rangle_{N-1}$ has an infinite number of terms as $\Gamma\rightarrow\infty$. However, there are terms of $\langle\sum_{m=1}^{N-1} H_{\mu_m}\rangle_{N-1}$ whose contributions are more important than others. For a general value of $N$ it is difficult to identify the most important contributions except for $N \leq 4$ where the most significant terms are generated by partitions $\mu$ close to the last partition $\Lambda$ given by
\[
\Lambda = \left\{ \begin{array}{rl}
 \left(\mu_0-1,\mu_0-2,\ldots,\mu_0-(N-1),0\right) &\mbox{ with $\mu_0=\frac{1}{4}N(\Gamma+2)$ if $\frac{\Gamma}{2}$ is even}  \\
  \left(\mu_0,\mu_0,\ldots,\mu_0,0\right) & \mbox{ with $\mu_0=\frac{1}{4}N\Gamma$ if $\frac{\Gamma}{2}$ is odd}
       \end{array} \right. .
\]
If the term generated by the last partition is the most important, then 
\[
\langle\sum_{m=1}^{N-1} H_{\mu_m}\rangle_{N-1} \underset{\Gamma \gg 1}{=} (N-1) H_{N\Gamma/4} \hspace{1.0cm} \mbox{for} \hspace{1.0cm}  N \leq 4 
\]
and the excess energy takes the form
\begin{equation}
 \left.U_{exc}\right|_{N=2,3,4} \underset{\Gamma\gg1}{=} \frac{N q^2}{4}\left\{(N-1)\left[H_{(N-1)\Gamma/2+1}-H_{N\Gamma/4}\right] + \log\left(\frac{N}{\rho_b \pi L^2}\right) - N \right\}.  
 \label{UexcInTheAsymptoticLimitNleq4Eq}
\end{equation}
Finally, in the limit $\Gamma\rightarrow\infty$ the energy of the Wigner crystal is
\begin{equation}
 \left. U_{exc} \right|_{N=2,3,4} \underset{\Gamma\rightarrow\infty}{=} \frac{N q^2}{4}\left\{(N-1)\log\left[2\left(1-\frac{1}{N}\right)\right]+\log\left(\frac{N}{\rho\pi L^2}\right)-1\right\}
 \label{UexcInTheInfinityNleq4Eq}
\end{equation}
both expressions Eq.~(\ref{UexcWignerNleq4Eq}) and Eq.~(\ref{UexcInTheInfinityNleq4Eq}) give the same result for $N \leq 4$. The asymptotic behaviour of the excess energy is shown in Fig. \ref{UexcInTheAsymptoticLimitNleq4Fig}. For $N>4$ Eq.~(\ref{UexcInTheInfinityNleq4Eq}) fails as it happens with $N=6$ where the corresponding geometrical factor of the octahedron is $\eta_6 = \frac{2}{\sqrt{2}}$  because it is necessary to include more terms in the computation of $\langle\sum_{m=1}^{N-1} H_{\mu_m}\rangle_{N-1}$ and eventually all of them for sufficiently large number of particles. 

\subsection{Energy for $N \geq 4$ with $\Gamma>2$ and comparison with
  the Metropolis method}
\label{subsec:energyNlarge}

\begin{figure}[h]
\includegraphics[width=0.33\linewidth]
{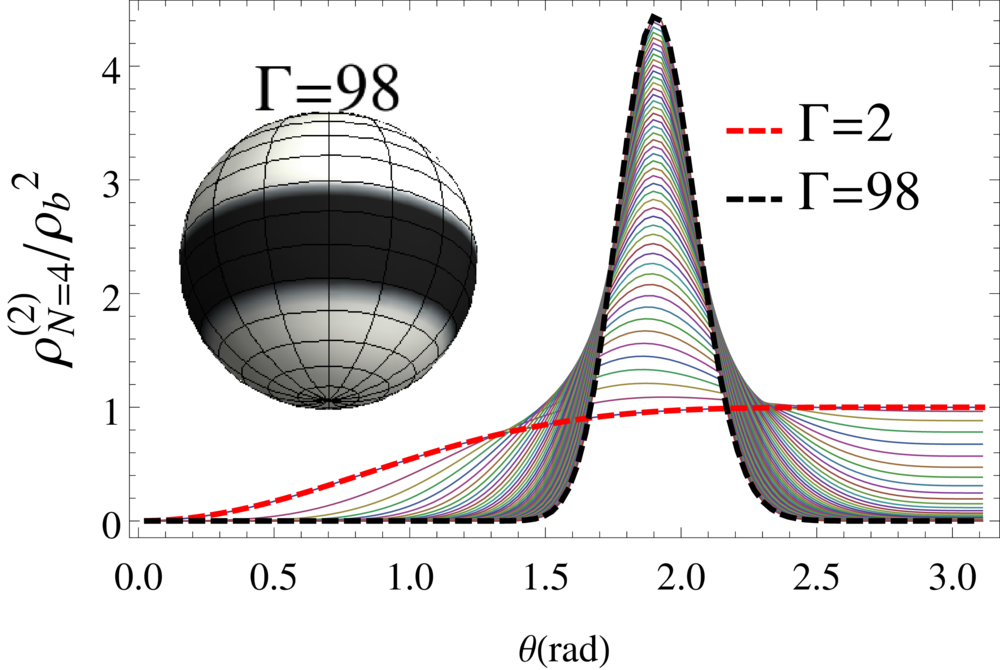}
\includegraphics[width=0.33\linewidth]
{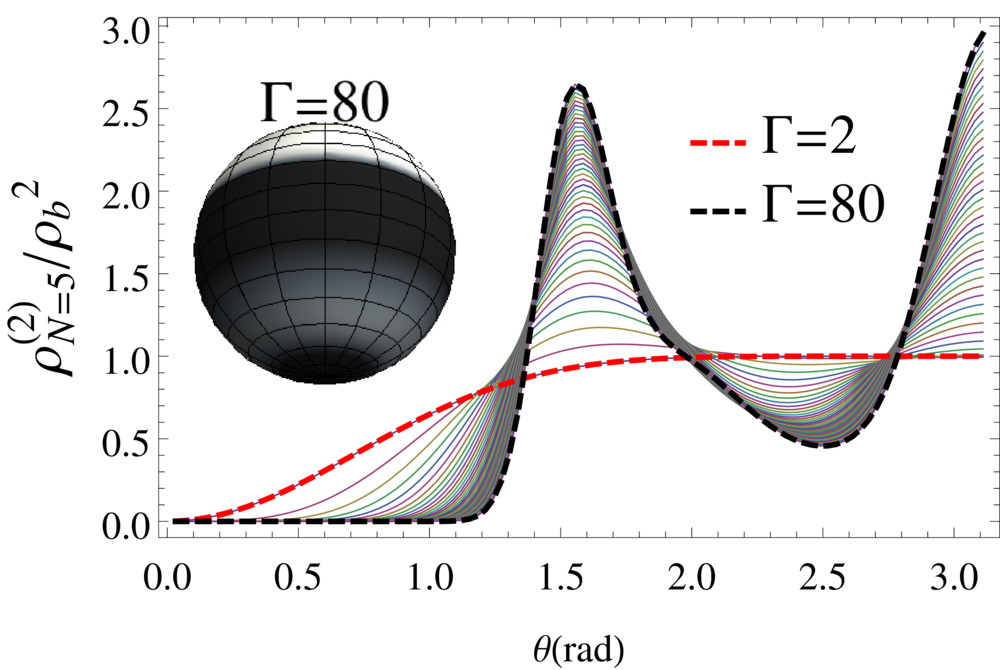}
\includegraphics[width=0.33\linewidth]
{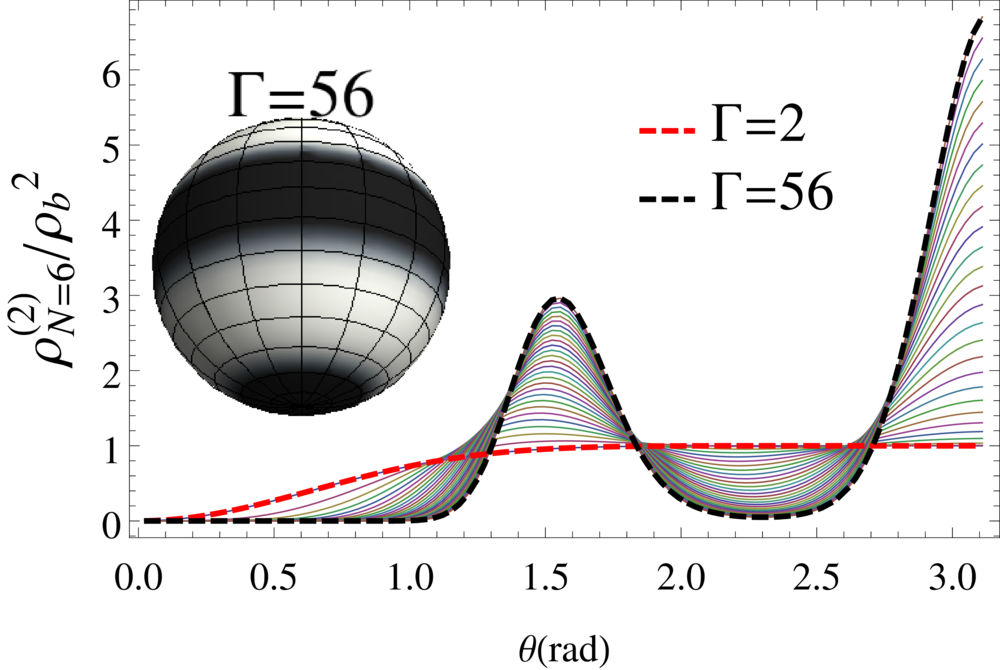}\\
\includegraphics[width=0.33\linewidth]
{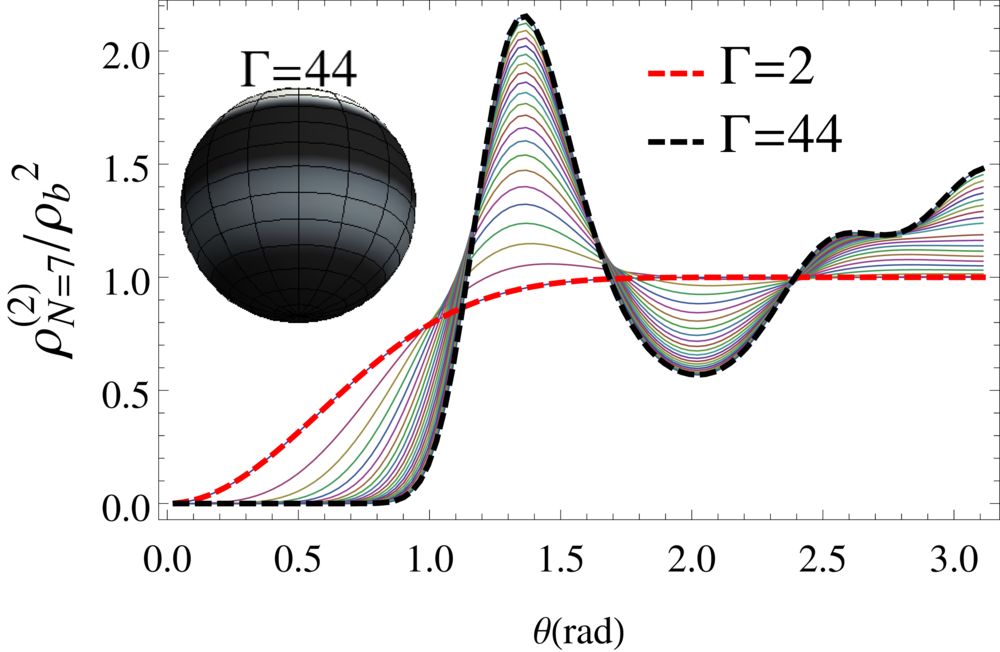}
\includegraphics[width=0.33\linewidth]
{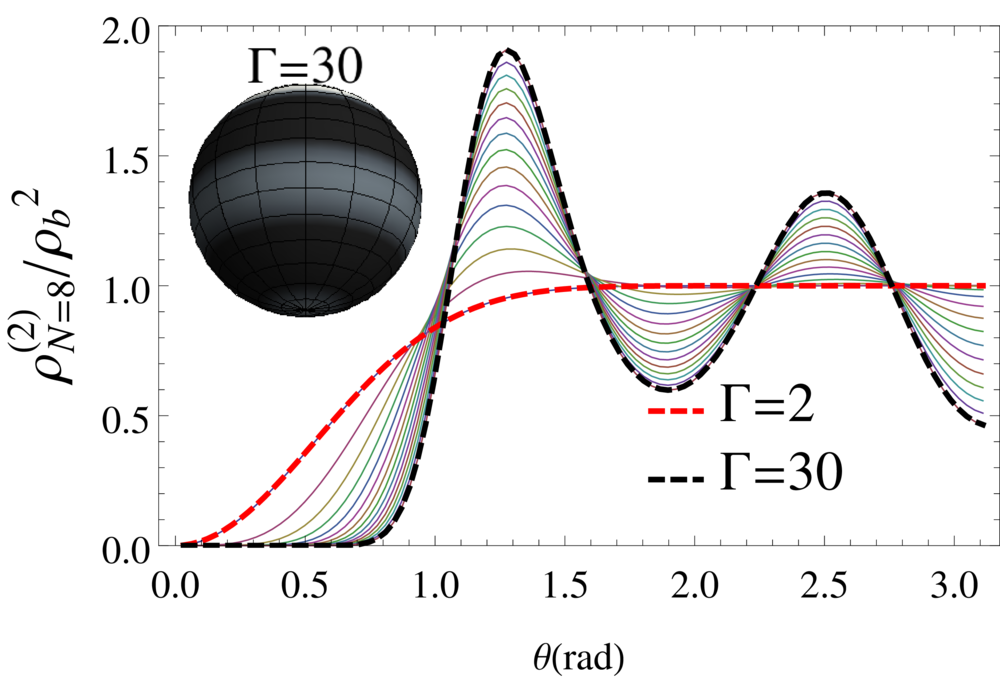}
\includegraphics[width=0.33\linewidth]
{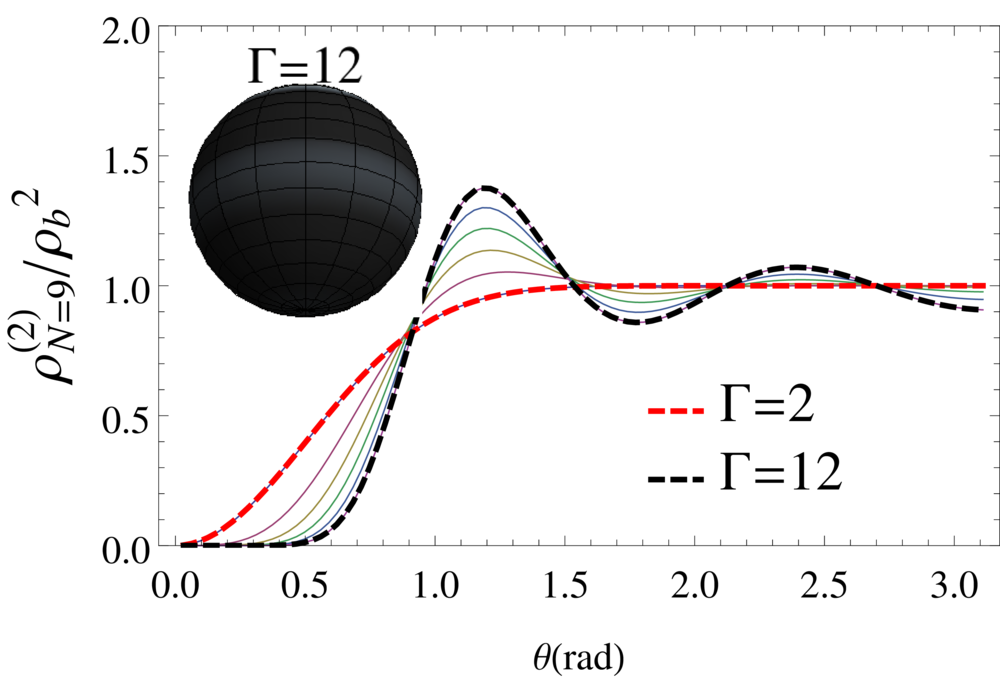}
\caption[Correlation function for $N=3$.]%
  {Numerical exact correlation function of the 2dOCP on the sphere for $N=4,5,\ldots,9$ particles.}  
\label{correlationFunctionForN4_5_6_7_8_9Fig}
\end{figure}

Previous works \cite{MontecarloStudyCaillol} showed that Monte Carlo simulations of 2dOCP are reliable for a wide range of $\Gamma\in\left[0.5,200\right]$. This section of the document is devoted to the implementation of the usual Metropolis method in the modest situation of a few particles 2dOCP with the aim to do a comparison between the Metropolis method and the exact results described in previous sections. In order to implement the numerical algorithm we chose randomly a particle located at $\vec{r}$ on the sphere, later the particle was moved to $\vec{r}' = \hat{R}_x(\gamma) \hat{R}_y(\beta) \hat{R}_z(\alpha) \vec{r}$ where $\hat{R}_{x}, \hat{R}_{x}$ and $\hat{R}_{y}$  are rotations around the $x, y$ and $z$ axis of the Cartesian reference frame in the center of the sphere. If the $k$-th particle is moved, then energy change is $\delta U_k =  -q^2\sum_{i=1, i\neq k}^N\log\left(\frac{|\vec{r}_i-\vec{r}'_k|}{|\vec{r}_i-\vec{r}_k|}\right).$
\begin{figure}[h]
  \centering   
  \includegraphics[width=0.9\textwidth]{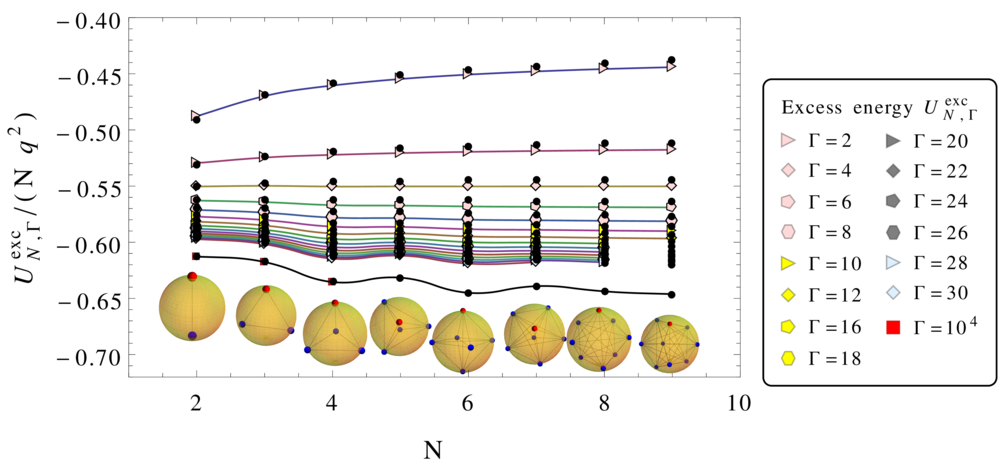}  
  \caption[Excess energy]%
  {Excess energy per particle. The exact values computed according to
    Eq. (\ref{exactUexcAnyGammaEq}) are  plotted with polygons, and
    black points correspond to Metropolis method. The analytical
    excess energy per particle for $\Gamma \rightarrow \infty$ was
    computed with Eq.~(\ref{UexcInTheInfinityNleq4Eq}) and represented
    with red squares. In this plot we have set $\rho_b=1$ and $L=1$.}
\label{ExcessEnergyNumericAndExactFig}  
\end{figure}
\begin{figure}[h]
  \centering   
  \includegraphics[width=0.7\textwidth]{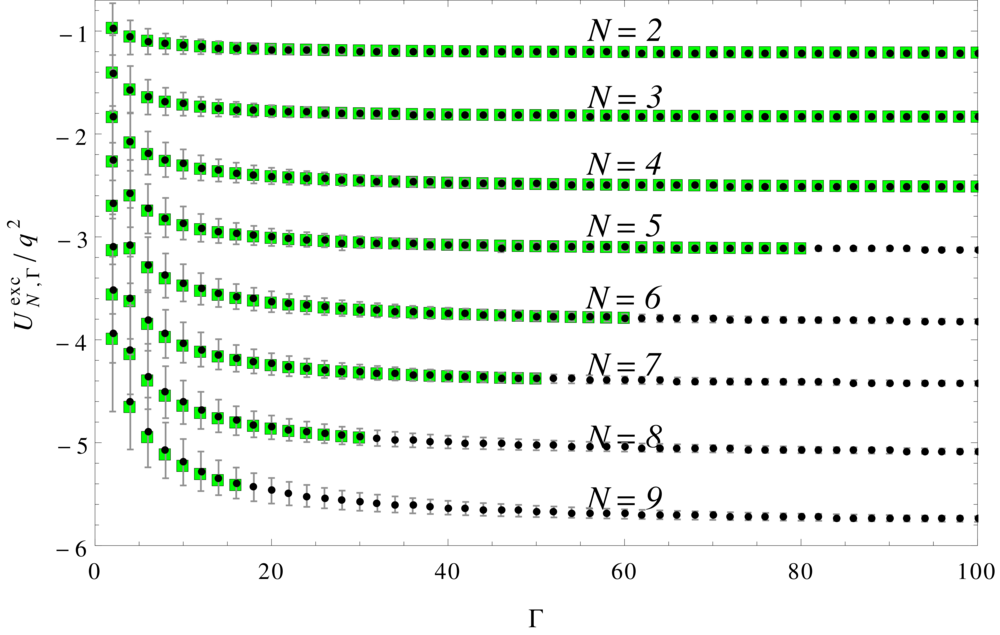}  
  \caption[Excess energy]%
  {Excess energy. Excess energy vs $\Gamma$ parameter. Black dots and gray error bars corresponds to Metropolis method. The exact numerical values are plot as squares.}
\label{ExcessEnergyNumericAndExact2Fig}  
\end{figure}
\begin{figure}[h]
  \centering   
  \includegraphics[width=0.13\textwidth]{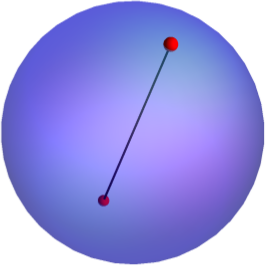}\hspace{0.5cm}
  \includegraphics[width=0.13\textwidth]{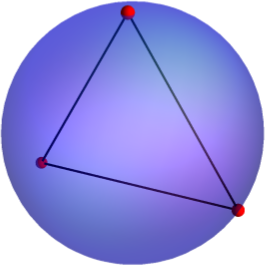}\hspace{0.5cm}
  \includegraphics[width=0.13\textwidth]{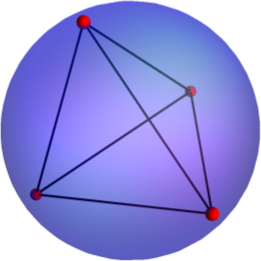}\hspace{0.5cm}
  \includegraphics[width=0.13\textwidth]{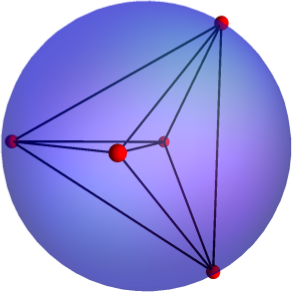}\\
  \includegraphics[width=0.13\textwidth]{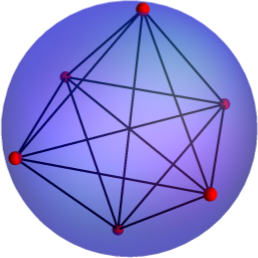}\hspace{0.5cm}
  \includegraphics[width=0.13\textwidth]{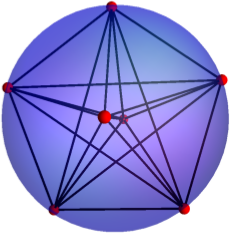}\hspace{0.5cm}
  \includegraphics[width=0.13\textwidth]{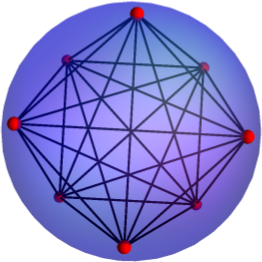}\hspace{0.5cm}
  \includegraphics[width=0.13\textwidth]{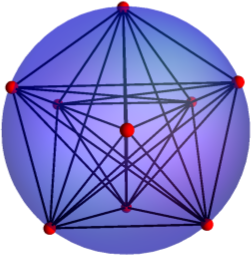}
  
  \caption[Configurations of the OdC2P on the sphere for large values of $\Gamma$.]%
  {Configurations of the OdC2P on the sphere for large values of $\Gamma$ and $N=2,3,\ldots 9$ particles. We have set a random initial condition and $\Gamma = 10^9$ in the Monte Carlo simulation.}
\label{coulombCrystalsFig}  
\end{figure}
\\As usual, the step is accepted if the energy of the new
configuration decreases or the Boltzmann factor
$\exp\left(-\beta\delta U_k\right)<M$ with $M\in[0,1]$ a random
number. The position of the mobile charges on the 2dOCP tends to be
fixed for large values of $\Gamma$ (this is high electric coupling or
low temperature) and the Metropolis method generates the
configurations shown in Fig. \ref{coulombCrystalsFig}. These
configurations are intimately related with the pair correlation
function (see Fig. \ref{correlationFunctionForN4_5_6_7_8_9Fig}) of
Eq.~(\ref{correlationFunctionEq}). For the case $N=4$ where particles
are arranged in the vertices of a tetrahedron and the pair correlation
function is concentrated around $\theta = 109.4712^o$ as $\Gamma$
reaches high values. For $N=5$ and $N=6$ the function $\rho_N^{(2)}$
is concentrated around $\theta = 180^{o}$ because there are different
ways to rotate the triangular dipyramid and the octahedron locating a
particle in the north pole and another one in the south pole. As we
add more particles new peaks on $\rho_N^{(2)}$ may emerge in the
strong coupling regime revealing the crystal structure of the
OCP. However, as $N$ is increased, it becomes less evident to
appreciate a direct connection between the individual positions of the
Wigner crystal and $\rho_N^{(2)}$ because there are more candidates to
occupy the north pole. We have also calculated the exact numerical
value of excess energy for $\Gamma=2,4,\ldots$ by using
Eqs. (\ref{excesEnergyEq}) and (\ref{exactUppEq}). The excess energy
tends to a constant as $\Gamma \gg 2$ as is shown in
Fig. \ref{ExcessEnergyNumericAndExact2Fig}. Such constant corresponds
to the minimal energy of the corresponding Wigner crystal. Similarly,
the excess energy per particle tends to a constant, say $u_{\Gamma}$,
as the number of particles is increased holding the density unchanged
(see Fig. \ref{ExcessEnergyNumericAndExactFig}). Although, the value
of $u_{\Gamma}$ has not been fully determined analytically at the
thermodynamic limit, it is interesting to notice that the excess
energy per particle at $\Gamma=2$ tends to
$u_{2}=-q^2\left[\log(\rho\pi L^2)+\gamma\right]/4$ according to
$U_{N,\Gamma=2}^{(exc)}/N = u_{2} - q^2/(8N) + O(1/N^2)$. In contrast,
energy per particle seems to oscillate around $u_{\Gamma}$ as $N$
increases for sufficiently large values of $\Gamma$.  \color{black}The
particle-particle energy computation via Eq.~(\ref{exactUppEq}) as
well as the pair correlation function
Eq.~(\ref{correlationFunctionEq}) require the knowledge of
$C^{(N)}_\mu(\Gamma/2)$ and partitions included in the sum. In
general, it may be computational expensive even for moderate values of
$N$ or $\Gamma$ because the number of coefficients tends to increase
quickly with these parameters. Fortunately, there are several
algorithm for the determination of the coefficients
\cite{computationalCoefficients1,computationalCoefficients2,computationalCoefficients3}
including the methods described in the Appendices~\ref{sec:app1} and
\ref{AppendixCoeff}. In particular, the approach of
\cite{computationalCoefficients3} seems to be the most efficient. The
results for the excess energy are shown in Tables
\ref{excessEnergyN2Table}-\ref{excessEnergyN9Table} of
Appendix~\ref{app:UexcGamma}, illustrating how $U_{exc}$ changes as
$\Gamma$ varies, for fixed values of the number of particles.

Alternatively, in Table~\ref{tab:UintG4_6_8} of
Appendix~\ref{app:UexcNfit}, it is shown how the excess energy varies
as $N$ increases for three fixed values of $\Gamma=4, 6, 8$. The data
shown in that table is computed with Eq. (\ref{exactUexcAnyGammaEq}) and fitted to an ansatz of the form
\begin{equation}
  \label{eq:Uansatz}
   U_{exc}=q^2(A N + B + C/N + D/N^2) 
\end{equation}
which allow us to estimate the bulk excess energy per particle
($u_{\Gamma}=A q^2$) and its finite size corrections. In particular,
we obtain $u_4/q^2=-0.5138290$, $u_6/q^2=-0.54990$ and
$u_8/q^2=-0.570$. We have fixed $L=1$ and $\rho_b=1$ in the numerical
data.

In Ref.~\cite{jancoviciFreeEnergyPrediction}
it is argued that the free energy $F$ of two-dimensional Coulomb
systems in a sphere is expected to have a finite-size expansion given by
\begin{equation}
  \label{eq:betaFlog}
  \beta F = N C_1 + \frac{\chi}{12} \ln N + C_2 + O(1/N)
\end{equation}
where $\chi=2$ is the Euler characteristic of the sphere. From the
relation $U=\partial (\beta F)/\partial \beta$ we deduce that the
expected finite size expansion of $U_{exc}$ should indeed be of the
form (\ref{eq:Uansatz}). Notice the absence of a logarithmic ($\ln N$)
finite-size correction in the internal energy $U$ as opposed to the
one appearing in the free energy (\ref{eq:betaFlog}), and in the
entropy, which will be discussed in the following section.

\section{Entropy}
\label{sec:entropy}
Using the definition of the Hemholtz free energy $F=U-TS$ the entropy is  
\[
S = \frac{U}{T} + k_B\log Z_c(T,A,N) = S_{ideal} + S_{exc}
\]
where $S_{ideal}$ is the entropy of the ideal bidimensional gas, and
$S_{exc}$ the excess entropy. Replacing the total energy
Eq.~(\ref{totalEnergy}) with $\langle U_{pp}\rangle$ and the configurational partittion function Eq.~(\ref{cofigurationalPartitionFunctionEq}), we find

\begin{equation}
S \underset{q=0}{=} S_{ideal} = N k_B\left[1+\log\left(\frac{A 2m\pi k_B T}{h^2}\right)\right]-k_B\log N!
\label{idealGasEntropyEq}
\end{equation}
and 
\[
S_{exc} =  N k_B\left\{ \frac{N\Gamma}{4} + \log\left[\left(\frac{L}{2R}\right)^{\Gamma/2}\right]\frac{1}{\left(\frac{\Gamma}{2}(N-1)+1\right)!}\right\} + k_B \log  Z^{sphere}_{N,\Gamma} + k_B\log N! +\frac{U_{exc}}{T}.
\]

\subsection{Entropy in the thermodynamic limit for $\Gamma=2$}
\begin{figure}[h]
  \centering   
  \includegraphics[width=0.56\textwidth]{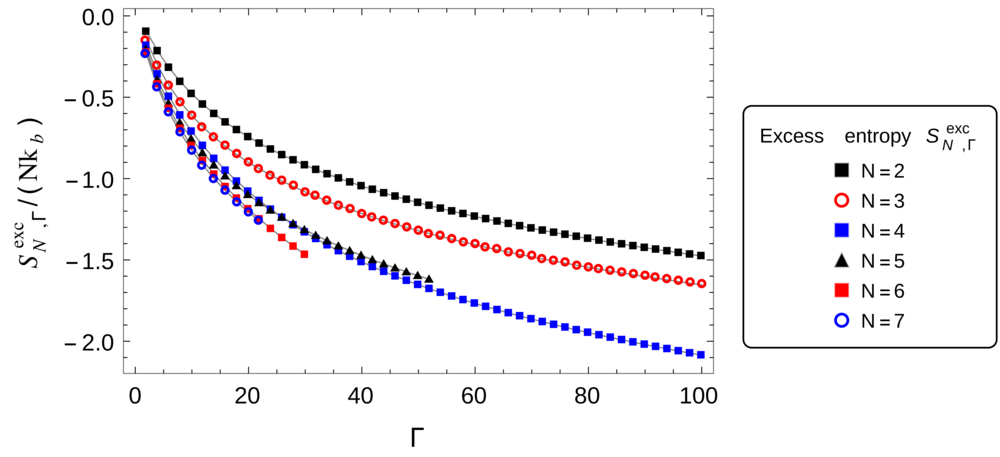}
  \caption[Entropy.]%
  {Excess entropy per particle for $\Gamma \geq 2$.  Numerical exact values of $S^{exc}_{N,\Gamma}/(N k_B)$ are represented by symbols for different values of $\Gamma$. The solid line is only a guide of the eye.}
\label{entropyForAnyGammaFig}  
\end{figure}

Using Eq.~(\ref{zSphereGamma2Eq}) we may write
\[
\log Z^{sphere}_{N,\Gamma} = 2 \log G_B(1+N)
\]
where $G_B(x)$ is the Barnes G-function. For the asymptotic limit
$N\to\infty$, we know  
\[
\log G_B(1+N) = \zeta'(-1) + \frac{N}{2}\log(2\pi)+\frac{1}{2}\left(N^2-\frac{1}{6}\right)\log N -3\frac{N^2}{4}+\sum_{k=1}^M\frac{B_{2k+2}}{4k(k+1)N^{2k}}+O\left(\frac{1}{N^{2M+2}}\right) 
\]
Where $\zeta(x)$ is the Riemann zeta function. Using this result with
the Stirling formula then the free energy, when the number of
particles $N\to\infty$, takes the form \cite{TellezForrester1999} 

\[
-\beta F = \log Z_{N,\Gamma=2}=\frac{N}{2}\log\left(\frac{2\pi^2 L^2}{\rho_b}\right)-\frac{\chi}{12}\log(N)+\left[2\zeta'(-1)-\frac{1}{12}\right]-\frac{1}{180 N^2}+O\left(\frac{1}{N^4}\right).
\]
This expression for the free energy also coincides with the one found by Jancovici et al. \cite{jancoviciFreeEnergyPrediction} where $F$ as a function of the number of particles is
\[
\beta F = C_1 N  + \frac{\chi}{12}\log(N)+ C_2+\cdots,
\]
where $C_1$ and $C_2$ are constants. On the other hand, the term
$\frac{U_{exc}}{T}$ in the limit $N\to\infty$, for $\Gamma=\frac{q^2}{k_B T}=2$, according to Eq.~(\ref{UexcForGamma2Eq}) is 
\begin{equation}
\frac{U_{exc}}{T} = -\frac{Nk_B}{2}\left[\log(\rho_b\pi L^2)+\gamma+\frac{1}{2N}-\frac{1}{12N^2}+\frac{1}{120N^4}+O\left(\frac{1}{N^6}\right)\right].
\end{equation}
As a result, the entropy, when $N\to\infty$, may be written as follows
\begin{multline}
S = \left\{1+\log\left(\frac{2m\pi k_B T}{\rho_b h^2}\right)+\frac{1}{2}\left[\log(2\pi)-\gamma\right]\right\}N k_B - k_B\frac{\chi}{12}\log N + \\ + k_B\left[2\zeta'(-1)-\frac{1}{3}\right]+k_B\left[\frac{1}{24N}-\frac{1}{180N^2}-\frac{1}{240N^3}\right]+O\left(\frac{1}{N^4}\right)
\end{multline}
where the excess entropy is 
\begin{multline}
S_{exc} = k_B\left\{\left[\frac{1}{2}\log(2\pi)-\frac{1}{2}\gamma-1\right]N + \frac{1}{3}\log N + \left[2\zeta'(-1)-\frac{1}{3}+\frac{1}{2}\log(2\pi)\right] + \right. \\ \left. + \left[\frac{1}{8N}-\frac{1}{180N^2}-\frac{1}{144N^3}\right]\right\}+O\left(\frac{1}{N^4}\right).
\end{multline}
\begin{figure}[h]
  \centering   
  \includegraphics[width=0.43\textwidth]{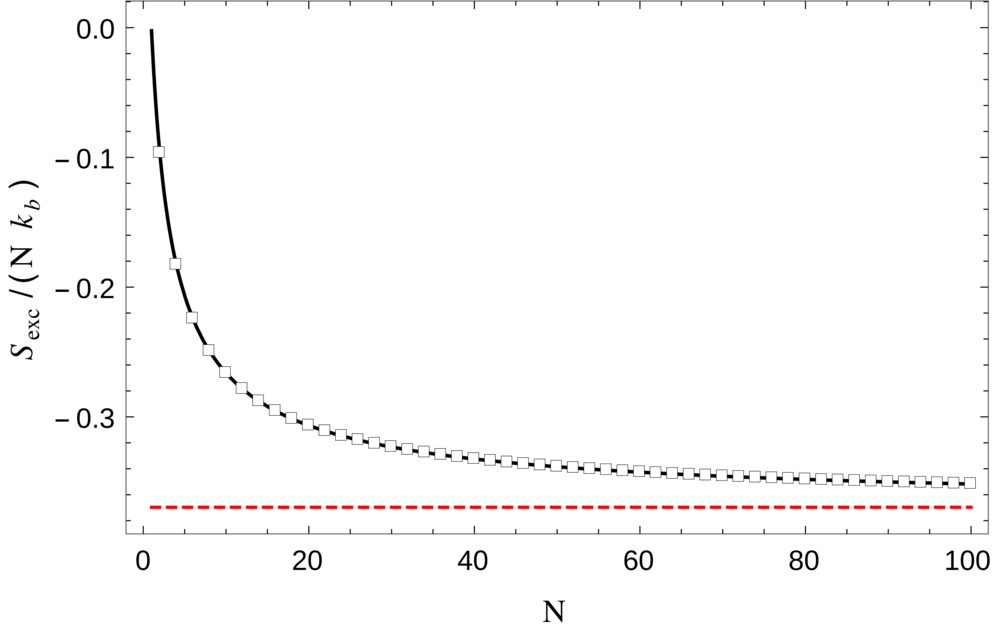}\hspace{0.5cm}
  \includegraphics[width=0.425\textwidth]{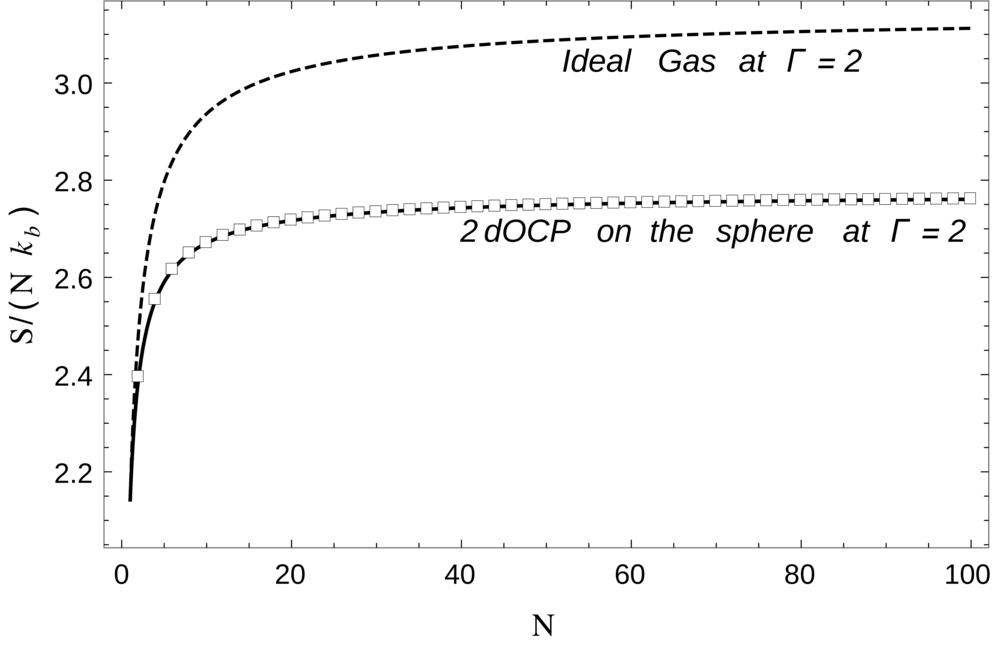}
  \caption[Entropy.]%
  {Entropy. \textbf{(left)} Excess entropy per particle. \textbf{(right)} Total entropy of the 2OCP for $\Gamma=2$. The asymptotic function of $S/N$ \textit{(solid line)} fits with the exact values \textit{(squares)} even for small values of $N$. The entropy of the ideal gas $S_{ideal}$ in the sphere \textit{(dashed line)} is bigger than $S$ for $N>1$. }
\label{entropyFig}  
\end{figure}
\\As a result, the total entropy, as well as the Helmholtz free
energy, is a function of the form $S = B_1 N + k_B\frac{\chi}{12}\log
N + B_2 + \cdots$, when $N\to\infty$, where $B_1$ and $B_2$ are constants. Moreover, the excess entropy $S_{exc}$ is negative for $N>2$ and decreases linearly as $N\rightarrow\infty$ as it is shown in Fig. \ref{entropyFig}-(left). However, the total entropy remains positive but lower than the entropy of the ideal gas on the sphere $0 < S \leq S_{ideal}$. It may be attributed to the fact that several microstates look less probable as the gas is charged. For instance, the situation where particles are too close is not too probable in the ideal gas, but the same situation is less probable if the particles are equally charged. The number of accessible microstates should decrease when the Coulomb interaction is introduced because there is less freedom to choose a configuration holding the total energy unchanged. The numerical exact values of the excess entropy for $\Gamma \geq 2$ are shown in Fig. \ref{entropyForAnyGammaFig}. In the regime of large values of $\Gamma$ the ideal gas entropy as well as the excess entropy take negative values contradicting the third law of thermodynamics which establishes that entropy of a perfect crystal at absolute temperature is zero. This problem is inherited from the classical treatment assumed in the study of 2dOCP where the entropy diverges as  $T \rightarrow 0$ to minus infinity suggesting that quantum treatment is necessary to find its value correctly in the strong coupling regime.
   
\section*{Concluding remarks}
In this work an approach was described to compute exactly some
thermodynamic properties of the 2dOCP on the sphere for $\Gamma \geq
2$ by expanding the Bolztmann factor in terms of monomial functions basis. The excess energy was exactly computed for even values of $\Gamma$ and $N=2,\ldots, 9$ particles obtaining a good agreement with the numerical simulations implemented with the Metropolis Method. The analytical expression for the excess entropy for even values of $\Gamma$ was also obtained. 

A pair of problems must be faced before computing the excess energy for $N \geq 4$ with the approach described in this document. The first one is to find the coefficients $C_{\mu}^{(N)}(\Gamma/2)$ included in the harmonic numbers average Eq.~(\ref{averageSumHEq}) and the second one is due to the rapid growing of the total number of partitions as the number of particles or the coupling parameter is increased. These problems are related with the practical evaluation of the expansion but not with the analytical expressions because they do not have restrictions with how large or cold the system is. A solution to the first task was to find a numerical method which allows to obtain the coefficients exactly. For this aim, we developed two different techniques based in the multinomial theorem and the difference method. They are alternatives to the numerical methods with recursion relations between coefficients. The number of terms of the expansions for the energy and other thermodynamic variables grows specially with $N$ but significantly lower when $\Gamma$ is increased. This feature permit us to obtain numerical exact results far from $\Gamma=2$ at least for a few number of particles finding a stronger connection between the analytical pair correlation function and the structure of the Wigner crystals as the coupling parameter becomes larger. 

\section*{Acknowledgments}
Authors would like to thank to Peter J. Forrester, Jean M. Calliol and
Martial Mazars for their valuable comments and discussions. Authors
also would like to thank to Nicolas Regnault for kindly facilitate
computational tools used in some of our computations.  This work was
supported by ECOS NORD/COLCIENCIAS-MEN-ICETEX, the Programa de
Movilidad Doctoral (COLFUTURO-2014) and Fondo de Investigaciones,
Facultad de Ciencias, Universidad de los Andes, project ``Exact
results for the mean energy of 2d Dyson Gas at $\Gamma>2$'', 2016-1.

\appendix
\section{Partitions computation}
\label{sec:app1}
All partitions $\mu$ may be found from the first one usually called the root partition $\mu_{i}^{1} = (N-i)\Gamma/2$ where we have used the following notation for the elements $\mu_{i}^{\alpha}$ of the partition $\mu$ with $(i=1,\ldots,N)$, $(\alpha=1,\ldots,\mathcal{N}_p)$ and $\mathcal{N}_p(N,\Gamma)$ the total number of partitions. For $N=3$ and $\Gamma=4$ the partitions are shown in Table \ref{partitionsExampleTable}. 
\begin{table}[h]
\begin{center}
  \begin{tabular}{ | c | c | }
    \hline
    \hline
    $\alpha$ & $(\mu_1^\alpha, \mu_2^\alpha, \mu_3^\alpha)$ \\ \hline
    \hline
    \hline    
    $1$ & $(4, 2, 0)$  \\ \hline
    $2$ & $(4, 1, 1)$  \\ \hline
    $3$ & $(3, 3, 0)$  \\ \hline
    $4$ & $(3, 2, 1)$  \\ \hline
    $5$ & $(2, 2, 2)$  \\         
    \hline
  \end{tabular}
  \caption{Partitions for $N=3$ and $\Gamma=4$.}
  \label{partitionsExampleTable}
\end{center}
\end{table}
\\The partition elements are obtained by adding or subtracting integers to the previous partition elements holding sum $\sum_{i=1}^N{\mu_i^\alpha}$ as a constant. For instance, the second partition $(4, 1, 1)$ may be obtained from $(4, 2, 0)$ by a subtraction of 1 from $\mu_2^1$ and adding 1 to $\mu_3^1$. Similarly, the third partition $(3, 3, 0)$ is obtained from the first partition by subtracting 1 from $\mu_1^1$ and adding 1 to $\mu_2^1$. These type of operations are usually referred as \textit{squeezing}. Following the same rules, the 4th partition may be obtained from the third and the 5th from the 4th. Therefore, the partition elements for $N=3$ are of the form $\mu_1^\alpha = (N-1)\Gamma/2-j_1(\alpha)$, $\mu_2^\alpha = (N-1)\Gamma/2+j_1(\alpha)-j_2(\alpha)$ and $\mu_3^\alpha = (N-1)\Gamma/2+j_2(\alpha)$ where $j_i(\alpha)$ are positive integers which represent the integers transferred from $\mu_i^\alpha$ to $\mu_{i+1}^\alpha$. In general, the partitions may be obtained from the following function  
\begin{equation}
f_i(N,\Gamma,\alpha) =
\left\{
	\begin{array}{ll}
		(N-i)\Gamma/2 - j_i  & \mbox{if  } i = 1\\
		(N-i)\Gamma/2 +j_{i-1} - j_i  & \mbox{if  } i \in [2, N)  \\
		(N-i)\Gamma/2 + j_{i-1} & \mbox{if  } i = N
	\end{array}
\right.
\label{functionForPartitionsEq}
\end{equation}
where $\mu_i^\alpha = f_i(N,\Gamma,\alpha)$ only if one of the following conditions is satisfied
\begin{equation}
\mu_1^\alpha \geq \mu_2^\alpha \geq \cdots \geq \mu_1^\alpha \mbox{ if } \Gamma/2 \mbox{ is even }
\label{partitionsEvenConditionEq}
\end{equation}
or
\begin{equation}
\mu_1^\alpha > \mu_2^\alpha > \cdots > \mu_1^\alpha \mbox{ if } \Gamma/2 \mbox{ is odd }
\label{partitionsOddConditionEq}
\end{equation}
The following lines of code written in Wolfram Mathematica 9.0 shows a
way to compute the partitions for $N=3$ particles.\\\\ 
\\$NConst=3;$ (*HOLD AS A CONSTANT IN THIS CODE*)\\
$\Gamma=8;$ (*GAMMA PARAMETER. $\Gamma/2$ MUST BE AN EVEN VALUE $\Gamma=4,8,12,\ldots$*)\\
$f[N_{-},\Gamma_{-},i_{-},JiMinus1_{-},Ji_{-}]:=Which[i==1,(N-i)\Gamma/2-Ji,(2\leq i \&\& i\leq N-1),(N-i)\Gamma/2 + JiMinus1-Ji, i==N, (N-i)\Gamma/2+JiMinus1];$ (*Eq.(\ref{functionForPartitionsEq})*)\\
$p=0$ (*PARTITIONS COUNTER*)\\
$For[j_1=0,j_1 \leq 2\Gamma,$\\
$\mbox{  }For[j_1=0,j_1 \leq 2\Gamma,$\\
$\mbox{    }f[i_{-}]:=f[NConst,\Gamma,i,j_{i-1},j_i];$\\
$\mbox{      }If[EvenQ[\Gamma/2]==True,$\\
$\mbox{        }If[f[1] \geq f[2] \&\& f[2] \geq f[3], p++; \mu[p,1]=f[1]; \mu[p,2]=f[2];\mu[p,3]=f[3]],$ (*EQ (\ref{partitionsEvenConditionEq})*)\\
$\mbox{        }If[f[1] > f[2] \&\& f[2] > f[3], p++; \mu[p,1]=f[1]; \mu[p,2]=f[2];\mu[p,3]=f[3]];];$ (*EQ (\ref{partitionsOddConditionEq})*)\\
$\mbox{   }j_2++];$\\
$j_1++];$\\\\
$NPartitions = p;$ (*NUMBER OF PARTITIONS*)\\
$Print[\mbox{``}There\mbox{ }are ",NPartitions,\mbox{'' }partitions\mbox{ }for\mbox{ }N = ",NConst, \mbox{``} and\mbox{ }\Gamma = ",\Gamma]$\\
$For[p=1, p\leq NPartitions, Print[\mbox{``}\mu=", Table[\mu[p,i],\left\{i,1,NConst\right\}]]; p++];$

\section{Coefficients computation} \label{AppendixCoeff}
\subsection{The multinomial theorem approach}
The coefficients may be computed from the following formula 
\begin{equation}
C_\mu^{(N)}(\Gamma/2) = \frac{1}{(2\pi)^N} \int_0^{2\pi}d\phi_1 e^{-\textit{\textbf{i}} \mu_1\theta_1}\cdots\int_0^{2\pi}d\phi_N e^{-\textit{\textbf{i}} \mu_N\theta_N} \prod_{1 \leq j<k \leq N} \left(e^{\textit{\textbf{i}}\theta_k}-e^{\textit{\textbf{i}}\theta_j}\right)^{\Gamma/2}
\label{coeffHalfGammaEven}
\end{equation}
\\with $\Gamma/2$ and even number. The product $\prod_{1 \leq j<k \leq N} \left(e^{\textit{\textbf{i}}\theta_k}-e^{\textit{\textbf{i}}\theta_j}\right)$ into the integral is a Vandermonde determinant 

\[
\Pi:=\prod_{1 \leq j<k \leq N} \left(e^{\textit{\textbf{i}}\theta_k}-e^{\textit{\textbf{i}}\theta_j}\right) = \sum_{\sigma\in S_N} \mbox{sgn}(\sigma)\prod_{j=1}^N \left(e^{\textit{\textbf{i}}\theta_j}\right)^{\sigma(j)-1} = \sum_{p=1}^{N!} \chi_{\sigma_p}  
\]
\\this is a sum of $N!$ terms of the form $\chi_p := \mbox{sgn}(\sigma^p)\exp\left[\textit{\textbf{i}}\sum_{i=1}^N \left(\sigma^p_i-1\right)\theta_i \right]$ with $S_N = \left\{\sigma^1,\ldots,\sigma^{N!}\right\}$ and $\sigma^p=\left\{\sigma^p_{1},\ldots,\sigma^p_{N}\right\}$ is the p-th permutation of $N$ elements. It is possible to use the multinomial theorem     
\[
\left(\sum_{p=1}^{M} \chi_p \right)^n = \sum_{i_1=0}^n \sum_{i_2=0}^{i_1} \cdots \sum_{i_{M-1} =0}^{i_{M-2}} \binom{n}{i_1} \binom{i_1}{i_2} \cdots \binom{i_{M-2}}{i_{M-1}} \chi_1^{n-i_1} \chi_2^{i_1-i_2} \cdots \chi_M^{i_{M-1}}
\]   
\\in order to evaluate the integral Eq.~(\ref{coeffHalfGammaEven}) where $n$ and $M$ are positive integers. If $M=N!$ and $n=\Gamma/2$ then
\begin{equation}
\Pi^{\Gamma/2} = \sum_{i_1=0}^{\Gamma/2} \sum_{i_2=0}^{i_1} \cdots \sum_{i_{N!-1} =0}^{i_{N!-2}} \binom{n}{i_1} \binom{i_1}{i_2} \cdots \binom{i_{N!-2}}{i_{N!-1}} \prod_{p=2}^{N!}\mbox{sgn}(\sigma^p)^{i_{p-1}-i_{p}}\exp\left( \textbf{\textit{i}}  \sum_{j=1}^N K_j(\vec{i};{\sigma_j})\theta_j\right)
\label{auxForCoeffEq}
\end{equation}
\\where we have defined
\begin{equation}
  K_m(\vec{i};{\sigma_m}) = K_m(i_1,\ldots,i_{N!-1}; \sigma_m ) := \sum_{j=1}^{N!}\left(i_{j-1}-i_j\right)\left(\sigma^j_m-1\right)
\label{KVectorEq}  
\end{equation}  
with $i_0:=\Gamma/2$, $i_{N!} := 0$, $\sigma^p_m$ is the $m$-th term of the permutation $\sigma^p$ and $\sigma_m=\left\{\sigma^1_m,\ldots,\sigma^{N!}_m\right\}$. For instance, the notation used for permutations of $N=3$ particles is 

\begin{equation}
\left(\sigma^p_m\right) = 
\left(\begin{matrix}
1	&2	&3\\
1	&3	&2\\
2	&1	&3\\
2	&3	&1\\
3	&1	&2\\
3	&2	&1  
\end{matrix}\right) = 
\left(\begin{matrix}
\sigma^1_1	&\sigma^1_2	 &\sigma^1_3 \\
\sigma^2_1	&\sigma^2_2	 &\sigma^2_3 \\
\sigma^3_1	&\sigma^3_2	 &\sigma^3_3 \\
\sigma^4_1	&\sigma^4_2	 &\sigma^4_3 \\
\sigma^5_1	&\sigma^5_2	 &\sigma^5_3 \\
\sigma^6_1	&\sigma^6_2	 &\sigma^6_3 \\
\end{matrix}\right)
\label{permutationMatrixEq}
\end{equation}
\\It is important to note that $K_m(\vec{i};{\sigma_m})$ is a non-negative integer because $i_j\leq i_{j-1}$ for $j=1,\ldots,N!$ and $m=1,\ldots,N$ as happens with the partition elements $\mu = \left\{\mu_m\right\} = (\mu_1,\ldots,\mu_N)$. Replacing Eq.~(\ref{auxForCoeffEq}) in Eq.~(\ref{coeffHalfGammaEven}) we obtain
\begin{multline*}
C_\mu^{(N)}(\Gamma/2) = \frac{1}{(2\pi)^N} \sum_{i_1=0}^{\Gamma/2} \sum_{i_2=0}^{i_1} \cdots \sum_{i_{N!-1} =0}^{i_{N!-2}} \binom{n}{i_1} \binom{i_1}{i_2} \cdots \binom{i_{N!-2}}{i_{N!-1}} \\ \prod_{p=2}^{N!}\mbox{sgn}(\sigma^p)^{i_{p-1}-i_{p}} \prod_{j=1}^{N}\int_0^{2\pi}d\phi_j \exp\left\{-\textit{\textbf{i}} (K_j-\mu_j)\theta_j\right\}.
\end{multline*}
\\The integration problem is solved by using $\int_0^{2\pi}d\phi_j \exp\left\{-\textit{\textbf{i}} (K_j-\mu_j)\theta_j\right\} = 2\pi\delta_{K_j,\mu_j} $. If the first permutation is the identity $\sigma^1_m = m$, then $\mbox{sgn}(\sigma^1) = 1$ and it is possible to write the coefficients more compactly
\begin{equation}
C_\mu^{(N)}(\Gamma/2) = \prod_{j=1}^{N!-1} \sum_{i_j=0}^{i_{j-1}}\left\{ \prod_{p=1}^{N!}\mbox{sgn}(\sigma^p)^{i_{p-1}-i_{p}} \binom{i_{p-1}}{i_p} \prod_{l=1}^{N} \delta_{K_l,\mu_l}\right\}.
\label{coefficientsWithDeltaProductEq}
\end{equation}
\\Now, the sum $\sum_{\vec{i}} := \prod_{j=1}^{N!-1} \sum_{i_j=0}^{i_{j-1}}$ of Eq.~(\ref{coefficientsWithDeltaProductEq}) generates a set vectors of the form $\vec{i}=\left(i_1,\ldots,i_{N!-1}\right)$. At the same time, each of these indices vector $\vec{i}$  will generate a set of non-negative integer elements $K = \left\{K_m(i_1,\ldots,i_{N!-1};\sigma_m) : m = 1\ldots,N\right\}$. Finally, the Kronecker delta product $\prod_{l=1}^{N} \delta_{K_l,\mu_l}$ collects only the $K$ sets which are partitions $K=\mu$. The number of $\vec{i}$ vectors generated with $\sum_{\vec{i}}$ is

\[
M_{N,\Gamma} = \sum_{\vec{i}} 1 = \frac{1}{(N-1)!}\prod_{j=1}^{N!-1}(1+\Gamma/2)
\]
\\and these vectors belong to the following set
\[
J(N,\Gamma):=\left\{\vec{i}=(i_1,\ldots,i_{N!-1}) : i_j \in \{0\}\cup\mathbb{Z}^+ \mbox{ with } j=1,\ldots, N!-1 \wedge \Gamma/2 \geq i_1 \geq i_2 \geq \cdots \geq i_{N!-1} \geq 0 \right\}.
\]
\\Hence,
\[
C_\mu^{(N)}(\Gamma/2) = \sum_{\vec{i} \in J} c(\vec{i})\delta_{\mu,K(\vec{i},\sigma)} \mbox{ with } c(\vec{i}) = \prod_{\sigma=1}^{N!}\mbox{sgn}(\sigma^p)^{i_{p-1}-i_{p}} \binom{i_{p-1}}{i_p}.
\]
If $I \subset J$ is the set of vectors $\vec{i}$ which generates a partition $K=\mu$, then each partition has a set $I_\mu \subset I$ of vectors $\vec{i}$ defined by 

\begin{multline}
I_\mu :=\left\{\vec{i} : \sum_{j=1}^{N!-1}\left(i_{j-1}-i_j\right)\left(\sigma_{jm}-1\right) + i_{N!-1}\left(\sigma_{N!m}-1\right)-\mu_m=0\hspace{0.1cm} \right. \\ \left. \forall \hspace{0.1cm} m=1,\ldots,N \wedge \frac{\Gamma}{2} \geq i_1 \geq i_2 \geq \cdots \geq i_{N!-1} \geq 0 \right\}.
\label{ImuSetEq}
\end{multline}
such that 
\[I(N,\Gamma)=\bigcup_{\mu=1}^{N_p(N,\Gamma)} I_\mu \]
because a set of indices $i_1,\ldots,i_{N!-1}$ in $I$ will generate a single partition and there are not two repeated partitions. As a result, the coefficients for a given number of particles $N$ and value of gamma parameter $\Gamma$ may be computed with
\begin{equation}
C_\mu^{(N)}(\Gamma/2) = \sum_{ (i_1,\ldots,i_{N!-1}) \in I_\mu} \prod_{p=1}^{N!}\mbox{sgn}(\sigma^p)^{i_{p-1}-i_{p}} \binom{i_p}{i_{p-1}}.
\label{coefficientsFinalResultEq}
\end{equation}
\\The computation of coefficients with Eq.~(\ref{coefficientsFinalResultEq}) requires to find the set $I_{\mu}$ previously defined in Eq.~(\ref{ImuSetEq}). In other words, it is necessary to solve a set of $N$-equations of the form $K_m(\vec{i},\sigma_m)-\mu_m=0$ with $m=1,\ldots,N$ under $N!$-conditions of the form $i_{j-1}-i_j \geq 0$ with $j=1,\dots,N!$, $i_0=\Gamma/2$ and $i_{N!}=0$ where $\vec{i}=(i_1,\ldots,i_{N!-1})$ are unknowns. Therefore, the vector solution is not unique $\mbox{dim}(I_\mu) \geq 1$ and it is possible to find more than one vector $\vec{i}$ solution associated to a single partition $\mu$ which makes harder to find the set $I_{\mu}$. The next code in addition with the one written in the previous section for partitions computation is an example of the coefficients computation for $N=3$ using Eq.~(\ref{coefficientsFinalResultEq}):\\\\   
$cb[n_{-},i_{-}]:=n!/((n-i)!i!); \mbox{ (*BINOMIAL COEFFICIENT*)} $\\
$\sigma List = Permutations[Table[i,\left\{i,1,NConst\right\}]];$\\ 
$\sigma[permutation_{-},i_{-}] := \sigma List[[permutation,i]]; \mbox{ (*PERMUTATIONS MATRIX EQ (\ref{permutationMatrixEq})*)} $\\
$i_0=\Gamma/2; i_{NConst!} = 0; \mbox{ (*FIRST AND LAST INDICES OF i-VECTOR*)}$\\\\
$vec=Table[FullSimplify[\sum_{m=1}^{NConst!}(i_{m-1}-i_m)(\sigma[m,j]-1),\left\{j,1,NConst\right\}];$\\
$K[j_{-}]:=vec[[j]];\mbox{ (*jth-COMPONENT OF K-VECTOR DEFINED IN EQ (\ref{KVectorEq})*)}$\\\\
$ISet[\mu 1_{-},\mu 2_{-},\mu 3_{-}]:=($\\
$solution=Solve[\Gamma/2 \geq i_1 \geq i_2 \geq i_3 \geq i_4 \geq i_5 \geq 0 \mbox{ \&\& } K[1]==\mu 1 \mbox{ \&\& } K[2]==\mu 2 \mbox{ \&\& } K[3]==\mu 3 , {i_1,i_2,i_3,i_4,i_5}, Integers ]; Return[\left\{ i_1,i_2,i_3,i_4,i_5 \right\}]/.solution);$ (*EQ (\ref{ImuSetEq})*) \\\\
$\mbox{(*COEFFICIENT FUNCTION*)}$\\
$coefficient[p_{-}]:=(set=ISet[\mu[p,1],\mu[p,2],\mu[p,3]] \mbox{ (*FIND THE SET } I_{\mu} \mbox{*) }$\\
$ISize=Dimensions[set][[1]]; \mbox{(*DIMENSION OF } I_\mu \mbox{*)}$\\
$iVec[n_{-},j_{-}]:=Wich[j==0,\Gamma/2,(1\leq j \&\& j\leq (NConst!-1)),set[[n]][[j]],j==NConst!,0]; \mbox{ (*}\vec{i} \in I_\mu \mbox{*)} $\\ 
$Print[\mbox{``} I_\mu=",MatrixForm[set]]; \mbox{ (*PRINT THE SET } I_\mu \mbox{*)} $\\
$c=\sum_{n=1}^{ISize}\prod_{j=1}^{NConst!}Signature[sigmaList[[j]]]^{iVec[n,j-1]-iVec[n,j]}cb[iVec[n,j-1],iVec[n,j]];$(*EQ (\ref{coefficientsFinalResultEq})*)\\
$Return[c];\mbox{(*RETURN THE COEFFICIENT*)});$\\\\
$For[p=1,p \leq NPartitions, Print[\mbox{``}\mu=",Table[\mu[p,i],\left\{i,1,NConst\right\}],\mbox{``} C=",coefficient[p]];p++];$
\\\\Here the set $I_\mu$ is found with the function \texttt{Solve} of Mathematica and the output is summarized in Table \ref{coefficientsExampleTable}. Taking into account that we usually have to compute the coefficients of partitions with $\mu_N=0$, then  the technique described here may be used for the computation of the coefficients for values of $\Gamma$ far from 2. In fact, we have used Eq.~(\ref{coefficientsFinalResultEq}) to find $C^{(N)}_\mu(\Gamma/2)$ for $N=2,3,4$ and $5$ with values of $\Gamma=4,8,\ldots,100$. In some sense the program implementation of Eq.~(\ref{coefficientsFinalResultEq}) may be simple, since the only difficulty is to built $I_\mu$. However, the construction of $I_\mu$ for large values of $N$ is hard even numerically because it will require to handle a set of $N!$ inequalities with $N$ equations. Although, this feature makes impractical the use of this technique even for $N > 5$, Eq.~(\ref{coefficientsFinalResultEq}) provides a straightforward way to find analytically $C^{(N)}_\mu(\Gamma/2)$ for $N=2$ and $N=3$ for any even value of $\Gamma$ when $\mu_N=0$.  
\begin{table}[h]
\begin{center}
  \begin{tabular}{ |c | c | c | c | }
    \hline
    \hline
    $\mu=(\mu_1,\mu_2,\mu_3)$ & $\vec{i}=(i_1,i_2,i_3,i_4,i_5)$ & $\mbox{dim}(I_\mu)$ & $C_\mu^{(N)}(\Gamma/2)$ \\ \hline
    \hline
    \hline    
    $(8,4,0)$ & $(4,4,4,4,4)$ & 1 & $1$ \\ \hline
    $(8,3,1)$ & $(4,4,4,4,3)$ & 1 & $-4$ \\ \hline
    $(8,2,2)$ & $(4,4,4,4,2)$ & 1 & $6$ \\ \hline
    $(7,5,0)$  & $(4,4,4,3,3)$ & 1 & $-4$ \\ \hline
    $(7,4,1)$  & $(4,4,4,3,2)$ & 1 & $12$ \\ \hline
    $(7,3,2)$  & $(4,4,3,3,3), (4,4,4,3,1)$ & 2 & $-8$ \\ \hline     
    $(6,6,0)$  & $(4,4,4,2,2)$ & 1 & $6$ \\ \hline
	$(6,5,1)	$  & $(4,3,3,3,3), (4,4,4,2,1)$ & 2 & $-8$ \\ \hline    
    $		$  & $(3,3,3,3,3), (4,3,3,3,2)$ & $ $ & $ $ \\
    $(6,4,2)	$  & $(4,4,3,2,2)$ & $4$ & $-22$ \\
    $		$  & $(4,4,4,2,0)$ & $ $ & $ $ \\ \hline
    $		$  & $(3,3,3,3,2)$ & $ $ & $ $ \\
    $(6,3,3)	$  & $(4,3,3,3,1)$ & $3$ & $48$ \\
    $		$  & $(4,4,3,2,1)$ & $ $ & $ $ \\ \hline
    $		$  & $(3,3,3,2,2)$ & $ $ & $ $ \\
    $(5,5,2)	$  & $(4,3,3,2,1)$ & $3$ & $48$ \\
    $		$  & $(4,4,3,1,1)$ & $ $ & $ $ \\ \hline
    $		$  & $(3,3,3,2,1), (4,3,2,2,2)$ & $$ & $$ \\
    $(5,4,3)	$  & $(4,3,3,2,0)$ & $4$ & $-36$ \\
    $		$  & $(4,4,3,1,0)$ & $ $ & $ $ \\ \hline
    $		$  & $(2,2,2,2,2), (3,2,2,2,1)$ & $ $ & $ $ \\
    $(4,4,4)	$  & $(3,3,2,1,1), (4,2,2,2,0)$ & $6$ & $90$ \\
    $		$  & $(4,3,2,1,0), (4,4,2,0,0)$ & $ $ & $ $ \\
    \hline
  \end{tabular}
  \caption{Coefficients for $N=3$ and $\Gamma=8$.}
  \label{coefficientsExampleTable}
\end{center}
\end{table}
\subsection{Coefficients for $N=2$}
The partitions are given by
\[
\mu_1^{\alpha} = \frac{\Gamma}{4} + \alpha - 1 \hspace{0.5cm}\mbox{and}\hspace{0.5cm} \mu_2^{\alpha} = \frac{\Gamma}{4} - \alpha + 1
\]
\\with $\alpha=1,2,\ldots, \mathcal{N}_p$ and $\mathcal{N}_p=\Gamma/4+1$ the total number of partitions. For $N=2$ the vector $\vec{i}$ has only one component $i_1$ and the $I_\mu$ set given by Eq.~(\ref{ImuSetEq}) takes the form    
\[
I_\mu :=\left\{i_1 : K_1(i_1,\sigma_1)=\mu_1 \wedge K_2(i_1,\sigma_2)=\mu_2 \mbox{ with } \frac{\Gamma}{2} \geq i_1 \geq 0 \wedge \mu_1 \geq \mu_2 \geq 0 \right\} \mbox{ with } (\sigma^p_m)=
\left(\begin{matrix}
1	&2\\
2	&1	
\end{matrix}\right).
\]
\\The components of the $K$-vector are $K_1(i_1,\sigma_1)=i_1=\mu_1$ and $K_2(i_1,\sigma_2) =\Gamma/2 - i_1=\mu_2$. From both conditions it is obtained the solution $i_1 = \Gamma/4 + (\alpha-1) = \mu_1$. Therefore, $\mbox{dim}(I_\mu)=1$  and the sum $\sum_{ (i_1,\ldots,i_{N!-1}) \in I_\mu}$ of Eq.~(\ref{coefficientsFinalResultEq}) has only one term  
\[
C_\mu^{(N)}(\Gamma/2) = \prod_{p=1}^{2}\mbox{sgn}(\sigma^p)^{i_{p-1}-i_{p}} \binom{i_p}{i_{p-1}}
\]
\\with $i_0=\Gamma/2$, $i_1=\mu_1$ and $i_2 = 0$. The final result is
\begin{equation}
C_\mu^{(2)}(\Gamma/2) = (-1)^{\mu_1} \binom{\Gamma/2}{\mu_1}.
\label{coefficientsForN2ResultEq}
\end{equation}  
This is the same formula obtained by combination of the binomial theorem and Eq.~(\ref{coeffHalfGammaEven}). 

\subsection{Coefficients for $N=3$ with $\mu_N=0$ } \label{AppendixCoeffN3}
The partitions and coefficients of $N$ are connected with the previous ones of $N-1$ according to the properties $\mu_j^{\alpha(N-1,\Gamma)} + \Gamma/2 = \mu_j^{\alpha(N,\Gamma)}$ and $C_\mu^{(N)}(\Gamma/2)=C_\mu^{(N-1)}(\Gamma/2)$ when $\mu_N = 0$. For this reason, the coefficients computation problem for $N=3$ particles with $\mu_3=0$ is similar to the problem for $N=2$ of the previous section. In this case the partitions are given by
\begin{equation}
\mu_1^{\alpha} = \Gamma + 1 - \alpha \mbox{,}\hspace{0.5cm} \mu_2^{\alpha} = \Gamma/2 + \alpha - 1 \hspace{0.5cm}\mbox{and}\hspace{0.5cm} \mu_3^{\alpha} = 0. 
\end{equation}
\\The number of partitions is found from the condition $\mu_1^{\mathcal{N}_p}=\mu_1^{\mathcal{N}_p}$ when $\Gamma/2$ is an even number and $\mu_1^{\mathcal{N}_p}+1=\mu_1^{\mathcal{N}_p}$ when $\Gamma/2$ is an odd number. The result is $\mathcal{N}_p=\mbox{Int}\left(\Gamma/4+1\right)$ and the set $I_\mu$ is 
\[
I_\mu :=\left\{\vec{i} : K_1(\vec{i},\sigma_1)=\mu_1 \wedge K_2(\vec{i},\sigma_2)=\mu_2 \wedge K_3(i_1,\sigma_3)=0 \mbox{ with } \frac{\Gamma}{2} \geq i_1 \geq i_2 \geq \ldots \geq i_5 \geq 0 \right\} 
\]
\\where $\vec{i}=\left(i_1,\ldots,i_5\right)$. The components of the $K$-vector are
\[K_1(\vec{i},\sigma_1) = i_2 + i_4 ,\]
\[K_2(\vec{i},\sigma_2) = \Gamma/2 + i_1 - 2i_2 + 2i_3 - 2i_4 + i_5 ,\]
\[K_3(\vec{i},\sigma_3) = \Gamma - i_1 + i_2 - 2i_3 + i_4 - i_5 ,\]
and the permutation matrix is given by Eq.~(\ref{permutationMatrixEq}). The solution is the following
\begin{equation}
i_1 = i_2 = i_3 = \frac{\Gamma}{2} \mbox{ and } i_4 = i_5 = \frac{\Gamma}{2} + 1 - \alpha.
\label{iVectorForN3Eq}
\end{equation}
\\Therefore $I_\mu$ has only one vector $\vec{i}$ when $\mu_3=0$ as is shown in Table \ref{coefficientsExampleTable} for the particular case of $\Gamma=8$ and Eq.~(\ref{coefficientsFinalResultEq}) takes the form
\[
C_\mu^{(N)}(\Gamma/2) = \prod_{p=1}^{6}\mbox{sgn}(\sigma^p)^{i_{p-1}-i_{p}} \binom{i_p}{i_{p-1}}
\]
\\with $i_0=\Gamma/2$ and $i_6 = 0$. The final result is
\[
C_\mu^{(3)}(\Gamma/2) = (-1)^{\mu_1} \binom{\Gamma/2}{\mu_2-\Gamma/2} \mbox{ with } \mu_3 = 0.
\]

\section{Exact coefficients computation via finite difference method (FDM)}
We know that Vandermonde determinant $ \det(z_j^{i-1})_{(i,j=1,2,\ldots,N)}$ to the power $\Gamma/2$ with $\Gamma$ a positive even number may be written in terms of the expansion 
\[
\Delta_N(z_1,\ldots,z_N)^{\Gamma/2} = \prod_{1 \leq i<j\leq N} \left(z_i-z_j\right)^{\Gamma/2} = \sum_{\mu}C_{\mu}^{(N)}(\Gamma/2)m_\mu(z_1,\ldots,z_N)
\]
\\where $\mu:=(\mu_1,\ldots,\mu_N)$ is a partition of $N(N-1)\Gamma/4$ with the condition $(N-1)\Gamma/2\geq\mu_1\geq\mu_2\cdots\geq\mu_N\geq 0$ where $m_\mu(z_1,\ldots,z_N)$ are the monomial symmetric functions  

\[
m_{\mu}(z_1,\ldots ,z_N) = \frac{1}{\prod_i m_i !} \sum_{\sigma\in S_N} z_{\sigma_1}^{\mu_1}\cdots z_{\sigma_N}^{\mu_N}
\]
when $\Gamma/2$ is even or antisymmetric functions when $\Gamma/2$ is odd. Since $\Delta_N(z_1,\ldots,z_N)^{\Gamma/2}$ is a polynomial with a finite number of terms whose exponents grouped in partitions $\mu$ determine completely each coefficient $C_{\mu}^{(N)}(\Gamma/2)$ for a given $N$ and $\Gamma$ then    

\[
\frac{\partial^{\mu_1 + \cdots + \mu_N } }{\partial z_1^{\mu_1} \cdots \delta z_1^{\mu_N}} \Delta_N(z_1,\ldots,z_N)^{\Gamma/2} = C_{\mu}^{(N)}(\Gamma/2) \frac{1}{\prod_i m_i !}  \sum_{\sigma\in S_N} \frac{\partial^{\mu_1 + \cdots + \mu_N } }{\partial z_1^{\mu_1} \cdots \delta z_1^{\mu_N}} \left( z_{\sigma_1}^{\mu_1}\cdots z_{\sigma_N}^{\mu_N} \right)  
\]
\\where

\[
\sum_{\sigma\in S_N} \frac{\partial^{\mu_1 + \cdots + \mu_N } }{\partial z_1^{\mu_1} \cdots \delta z_1^{\mu_N}} \left( z_{\sigma_1}^{\mu_1}\cdots z_{\sigma_N}^{\mu_N} \right) = \prod_i m_i ! \left( \mu_1! \cdots \mu_N!  \right)
\]
\\hence

\begin{equation}
C_{\mu}^{(N)}(\Gamma/2) = \frac{1}{\mu_1!\ldots\mu_N!} \frac{\partial^{\mu_1 + \cdots + \mu_N } }{\partial z_1^{\mu_1} \cdots \delta z_1^{\mu_N}} \left[ \Delta_N(z_1,\cdots,z_N)^{\Gamma/2} \right]
\label{coefficientsWithDerivativesEq}
\end{equation}
\\Finite difference method enable us to compute derivatives of functions approximately starting from the usual limit definition of derivatives
\[
\frac{d }{d x} f(x) = \lim_{h\rightarrow 0} \frac{f(x+h)-f(x)}{h} = \frac{\Delta_h^1 f(x)}{h} + O(h)
\]
\\with $\Delta_h^1 f(x) = f(x+h)-f(x)$ for \textit{forward difference}. The second derivative is 
\[
\frac{d }{d x} f(x) = \frac{\Delta_h^2 f(x)}{h^2} + O(h)
\]
where $\Delta_h^2 f(x) = \Delta_h^1( \Delta_h^1 f(x) ) = \Delta_h^1 f(x+h) - \Delta_h^1 f(x) = f(x + 2h) - 2f(x+h) + f(x)$. This procedure may be generalized in order to approximate the $n$-order derivative
\begin{equation}
\frac{d^n }{d x^n} f(x) = \frac{1}{h^n} \sum_{i=0}^{n} (-1)^i \binom{n}{i}f(x+(n-i)h) + O(h)
\label{firstDerivativeEq}
\end{equation}
\\For a general function $f(x)$ Eq.~(\ref{firstDerivativeEq}) give us an approximation, except in the particular case \textit{when $f(x)$ is a polynomial of order $n$ where Eq.~(\ref{firstDerivativeEq}) coincides with the exact result} by virtue of
\begin{equation}
\sum_{i=0}^{n} (-1)^i \binom{n}{i} (x+(n-i)h)^m = \left\{
	\begin{array}{ll}
		n!h^n  & \mbox{ if }  m=n \\
		0 & \mbox{ if } 0 \leq m < n \\
		\mbox{a function of } x \mbox{ and } h \mbox{ if } m > n
	\end{array}
\right.     
\label{derivativePropertyEq}
\end{equation}
\\The cases for which $0 \leq m \leq n$ are independent of the value of $x$ since they are cancelled in the expansion. As a result, we may write
\[
\frac{d^n }{d x^n} x^n = n! = \frac{1}{h^n} \sum_{i=0}^{n} (-1)^i \binom{n}{i} (x+(n-i)h)^n 
\]
\\Since $n!$ is a constant we may choose freely the value of $x$. If we set $x=0$, then
\[
\frac{d^n }{d x^n} x^n = \sum_{i=0}^{n} (-1)^i \binom{n}{i} (n-i)^n 
\]
\\Therefore, if $f(x) = \sum_{i=1}^n c_i x^i$ then
\[
\frac{d^n }{d x^n} f(x) = \sum_{i=0}^{n} (-1)^i \binom{n}{i} f(n-i)^n = n! c_n 
\]
\\and
\[
\frac{\partial^n }{\partial x^n}\frac{\partial^m }{\partial y^m} F(x,y) = \sum_{i=0}^{n} (-1)^i \binom{n}{i} f(n-i)^n \sum_{j=0}^{m} (-1)^j \binom{m}{j} g(m-j)^m = n! m! c_n p_m 
\]
where $F(x,y):=f(x)g(y)$ with $g(y) = \sum_{i=1}^m p_i y^i$ another polynomial of order $m$. More generally we may write

\begin{equation}
\frac{\partial^N }{\partial x_1^{n_1} \cdots \partial x_N^{n_N}} F(x_1,\ldots,x_N) = \sum_{i_1=0}^{n_1}\cdots\sum_{i_N=0}^{n_N} (-1)^{i_1 + \cdots + i_N} \binom{n_1}{i_1} \cdots\binom{n_N}{i_N} f( n_1-i_1, \ldots, n_N-i_N ) 
\label{nOrderDerivativeFDMEq}
\end{equation}
if $F(x_1,\ldots,x_N) = \prod_{i=1}^N f_i(x_i)$ with $f_i(x_i)$ a polynomial function of order $n_i$. If any function $f_i(x_i)$ of $F(x_1,\ldots,x_N)$ would have and order $n'_i$ lower than $n_i$ then Eq.~(\ref{nOrderDerivativeFDMEq}) would be simply zero because of Eq.~(\ref{derivativePropertyEq}) and would give you a wrong derivative if $n'_i > n_i$. Now consider the case 
\[
F(x_1,\ldots,x_N) = \sum_{i_1=0}^{n'_1}\cdots\sum_{i_N=0}^{n'_N} c_{i_1 \ldots i_N} x^{i_1}\ldots x^{i_N}
\]
where $i_1 + \ldots + i_N = constant$. We may obtain the coefficient $c_{i_1 \ldots i_N}$ applying $\frac{\partial^N }{\partial x_1^{i_1} \cdots \partial x_N^{i_N}}$ according to (\ref{nOrderDerivativeFDMEq}) even when any derivative $\frac{\partial }{\partial x_j^{i_j}}$ of another term say $c_{i'_1 \ldots i'_N} x^{i'_1}\ldots x^{i'_N}$ give us a wrong result if $i_j < i'_j$ because the restriction $i_1 + \ldots + i_N = constant$ ensures the existence of at least one derivative say $\frac{\partial }{\partial x_k^{i_k}}$ with $i_j > i'_j$ which transform the whole term in zero. This is just the case of Eq.~(\ref{coefficientsWithDerivativesEq}) because the partition elements have the constrain $\mu_1 + \ldots + \mu_N = N(N-1)\Gamma/4$ hence the coefficients for even values of $\Gamma/2$ take the form

\begin{equation}
C_{\mu}^{(N)}(\Gamma/2) = \frac{1}{\mu_1!\ldots\mu_N!}\sum_{i_1=0}^{\mu_1} \cdots \sum_{i_N=0}^{\mu_N} (-1)^{i_1 + \cdots + i_N} \binom{\mu_1}{i_1} \cdots\binom{\mu_N}{i_N} \left[ \Delta_N(\mu_1-i_1,\cdots,\mu_N-i_N)^{\Gamma/2}\right].
\label{cefficientFormulaFiniteDiffEq}
\end{equation}
In principle, the coefficients computation with Eq.~(\ref{cefficientFormulaFiniteDiffEq}) does not offer remarkable implementation difficulties. Nevertheless, it is important to note that $C_{\mu}^{(N)}(\Gamma/2)$ may have a large value as $\Gamma$ or $N$ increase. For instance, the coefficient for $N=5$ particles with  $\mu= (100,100,100,0)$ at $\Gamma=100$ is 
\[
C_{(100,100,100,0)}^{(N=5)}(40)=2042816020019820636556288572807323741663688000.\]
This value may easily overflow the maximum integer value permitted by the computer. Usually, this maximum value  varies with the program used to implement the coefficients computation formula as well as the architecture of the machine. Fortunately, in order to solve this problem it is possible to use multiple precision arithmetic libraries as GMP \cite{gmpLib} included in some of our computations.

\newpage
\section{Excess energy as a function of $\Gamma$}
\label{app:UexcGamma}
In this section we report the excess energy $U_{exc}$ obtained from
the exact expression, Eq.~(\ref{exactUexcAnyGammaEq}), the value obtained by Monte Carlo (MC)
simulations, and their relative difference. We have set $\rho_b=1$ and
$L=1$. The result presented in these tables is oriented to understand
how $U_{exc}$ depends on $\Gamma$, when it varies in a range from
$\Gamma=2$ up to a high coupling of $\Gamma=100$.

\begin{table}[!htb]    
    \begin{minipage}{.5\linewidth}
        \centering
        \begin{tabular}{ | c | c | c | c | }
    \hline
    \hline
    $\Gamma$ & $U_{exc}/q^2 \mbox{ Exact}$ & $U_{exc}/q^2 \mbox{ MC}$ & $\mbox{Error \%}$ \\ \hline
    \hline
    \hline    
    $2$	& $-0.9757913526$	& $-0.9812032812$	& $0.55$ \\ \hline
$4$	& $-1.059124686$	& $-1.062297677$	& $0.3$ \\ \hline
$6$	& $-1.100791353$	& $-1.101719325$	& $0.084$ \\ \hline
$8$	& $-1.125791353$	& $-1.124501433$	& $0.11$ \\ \hline
$10$	& $-1.142458019$	& $-1.138761235$	& $0.32$ \\ \hline
$12$	& $-1.154362781$	& $-1.152556455$	& $0.16$ \\ \hline
$14$	& $-1.163291353$	& $-1.160845284$	& $0.21$ \\ \hline
$16$	& $-1.170235797$	& $-1.16892675$	& $0.11$ \\ \hline
$18$	& $-1.175791353$	& $-1.173774336$	& $0.17$ \\ \hline
$20$	& $-1.180336807$	& $-1.17951004$	& $0.07$ \\ \hline
$22$	& $-1.184124686$	& $-1.182909902$	& $0.1$ \\ \hline
$24$	& $-1.187329814$	& $-1.18598303$	& $0.11$ \\ \hline
$26$	& $-1.190077067$	& $-1.188412893$	& $0.14$ \\ \hline
$28$	& $-1.192458019$	& $-1.192136479$	& $0.027$ \\ \hline
$30$	& $-1.194541353$	& $-1.193967292$	& $0.048$ \\ \hline
$32$	& $-1.196379588$	& $-1.195691052$	& $0.058$ \\ \hline
$34$	& $-1.198013575$	& $-1.197112376$	& $0.075$ \\ \hline
$36$	& $-1.199475563$	& $-1.199328565$	& $0.012$ \\ \hline
$38$	& $-1.200791353$	& $-1.200571597$	& $0.018$ \\ \hline
$40$	& $-1.201981829$	& $-1.201711017$	& $0.023$ \\ \hline
$42$	& $-1.20306408$	& $-1.202641379$	& $0.035$ \\ \hline
$44$	& $-1.204052222$	& $-1.203523393$	& $0.044$ \\ \hline
$46$	& $-1.204958019$	& $-1.204395568$	& $0.047$ \\ \hline
$48$	& $-1.205791353$	& $-1.205132481$	& $0.055$ \\ \hline
$50$	& $-1.206560583$	& $-1.205855639$	& $0.058$ \\ \hline

    \hline
  \end{tabular}
    \end{minipage}%
    \begin{minipage}{.5\linewidth}
      \centering
        \begin{tabular}{ | c | c | c | c | }
    \hline
    \hline
    $\Gamma$ & $U_{exc}/q^2 \mbox{ Exact}$ & $U_{exc}/q^2 \mbox{ MC}$ & $\mbox{Error \%}$ \\ \hline
    \hline
    \hline    
    $52$	& $-1.207272834$	& $-1.206448154$	& $0.068$ \\ \hline
$54$	& $-1.20793421$	& $-1.207096743$	& $0.069$ \\ \hline
$56$	& $-1.208549973$	& $-1.207607537$	& $0.078$ \\ \hline
$58$	& $-1.209124686$	& $-1.20811432$	& $0.084$ \\ \hline
$60$	& $-1.20966232$	& $-1.209532589$	& $0.011$ \\ \hline
$62$	& $-1.210166353$	& $-1.209999195$	& $0.014$ \\ \hline
$64$	& $-1.210639837$	& $-1.210423199$	& $0.018$ \\ \hline
$66$	& $-1.21108547$	& $-1.210878298$	& $0.017$ \\ \hline
$68$	& $-1.211505638$	& $-1.211198784$	& $0.025$ \\ \hline
$70$	& $-1.211902464$	& $-1.211574599$	& $0.027$ \\ \hline
$72$	& $-1.212277839$	& $-1.211914884$	& $0.03$ \\ \hline
$74$	& $-1.212633458$	& $-1.212267508$	& $0.03$ \\ \hline
$76$	& $-1.21297084$	& $-1.212531629$	& $0.036$ \\ \hline
$78$	& $-1.213291353$	& $-1.212839083$	& $0.037$ \\ \hline
$80$	& $-1.213596231$	& $-1.213093559$	& $0.041$ \\ \hline
$82$	& $-1.213886591$	& $-1.213368213$	& $0.043$ \\ \hline
$84$	& $-1.214163446$	& $-1.213651883$	& $0.042$ \\ \hline
$86$	& $-1.214427716$	& $-1.213884442$	& $0.045$ \\ \hline
$88$	& $-1.214680242$	& $-1.214087024$	& $0.049$ \\ \hline
$90$	& $-1.214921787$	& $-1.2143274$	& $0.049$ \\ \hline
$92$	& $-1.215153055$	& $-1.21451764$	& $0.052$ \\ \hline
$94$	& $-1.215374686$	& $-1.214730038$	& $0.053$ \\ \hline
$96$	& $-1.215587271$	& $-1.214933884$	& $0.054$ \\ \hline
$98$	& $-1.215791353$	& $-1.215103185$	& $0.057$ \\ \hline
$100$	& $-1.215987431$	& $-1.215285067$	& $0.058$ \\ \hline    
  \end{tabular}
    \end{minipage} 
    \caption{Excess energy for $N=2$.}
  \label{excessEnergyN2Table}
\end{table}
\newpage
\begin{table}[!htb]    
    \begin{minipage}{.5\linewidth}
        \centering
        \begin{tabular}{ | c | c | c | c | }
    \hline
    \hline
    $\Gamma$ & $U_{exc}/q^2 \mbox{ Exact}$ & $U_{exc}/q^2 \mbox{ MC}$ & $\mbox{Error \%}$ \\ \hline
    \hline
    \hline    
    $2$	& $-1.409588198$	& $-1.407061446$	& $0.18$ \\ \hline
$4$	& $-1.573873912$	& $-1.569847082$	& $0.26$ \\ \hline
$6$	& $-1.64946915$	& $-1.641879599$	& $0.46$ \\ \hline
$8$	& $-1.692884069$	& $-1.684435151$	& $0.5$ \\ \hline
$10$	& $-1.721049653$	& $-1.709615604$	& $0.66$ \\ \hline
$12$	& $-1.740796418$	& $-1.733258506$	& $0.43$ \\ \hline
$14$	& $-1.755406808$	& $-1.747236245$	& $0.47$ \\ \hline
$16$	& $-1.766653611$	& $-1.760380733$	& $0.36$ \\ \hline
$18$	& $-1.775578166$	& $-1.769152071$	& $0.36$ \\ \hline
$20$	& $-1.782832258$	& $-1.77840703$	& $0.25$ \\ \hline
$22$	& $-1.788844668$	& $-1.783992084$	& $0.27$ \\ \hline
$24$	& $-1.793909018$	& $-1.788840577$	& $0.28$ \\ \hline
$26$	& $-1.798233129$	& $-1.793231092$	& $0.28$ \\ \hline
$28$	& $-1.801968226$	& $-1.799063379$	& $0.16$ \\ \hline
$30$	& $-1.80522697$	& $-1.802011971$	& $0.18$ \\ \hline
$32$	& $-1.808095018$	& $-1.804868251$	& $0.18$ \\ \hline
$34$	& $-1.810638651$	& $-1.807262948$	& $0.19$ \\ \hline
$36$	& $-1.812909962$	& $-1.810750547$	& $0.12$ \\ \hline
$38$	& $-1.814950457$	& $-1.81269479$	& $0.12$ \\ \hline
$40$	& $-1.816793618$	& $-1.814506611$	& $0.13$ \\ \hline
$42$	& $-1.818466749$	& $-1.816214246$	& $0.12$ \\ \hline
$44$	& $-1.819992338$	& $-1.817602284$	& $0.13$ \\ \hline
$46$	& $-1.821389076$	& $-1.818976863$	& $0.13$ \\ \hline
$48$	& $-1.822672625$	& $-1.82010681$	& $0.14$ \\ \hline
$50$	& $-1.823856208$	& $-1.821293489$	& $0.14$ \\ \hline
  \end{tabular}
    \end{minipage}%
    \begin{minipage}{.5\linewidth}
      \centering
        \begin{tabular}{ | c | c | c | c | }
    \hline
    \hline
    $\Gamma$ & $U_{exc}/q^2 \mbox{ Exact}$ & $U_{exc}/q^2 \mbox{ MC}$ & $\mbox{Error \%}$ \\ \hline
    \hline
    \hline    
    $52$	& $-1.824951065$	& $-1.822395899$	& $0.14$ \\ \hline
$54$	& $-1.825966812$	& $-1.823401616$	& $0.14$ \\ \hline
$56$	& $-1.826911724$	& $-1.824288329$	& $0.14$ \\ \hline
$58$	& $-1.82779296$	& $-1.825153714$	& $0.14$ \\ \hline
$60$	& $-1.828616748$	& $-1.827197009$	& $0.078$ \\ \hline
$62$	& $-1.829388528$	& $-1.82795771$	& $0.078$ \\ \hline
$64$	& $-1.830113073$	& $-1.828654873$	& $0.08$ \\ \hline
$66$	& $-1.830794591$	& $-1.829326464$	& $0.08$ \\ \hline
$68$	& $-1.831436804$	& $-1.829845098$	& $0.087$ \\ \hline
$70$	& $-1.832043018$	& $-1.830498228$	& $0.084$ \\ \hline
$72$	& $-1.832616176$	& $-1.831067858$	& $0.084$ \\ \hline
$74$	& $-1.83315891$	& $-1.831593772$	& $0.085$ \\ \hline
$76$	& $-1.83367358$	& $-1.832024806$	& $0.09$ \\ \hline
$78$	& $-1.834162308$	& $-1.832566582$	& $0.087$ \\ \hline
$80$	& $-1.834627008$	& $-1.832970462$	& $0.09$ \\ \hline
$82$	& $-1.835069407$	& $-1.833419131$	& $0.09$ \\ \hline
$84$	& $-1.835491075$	& $-1.833823488$	& $0.091$ \\ \hline
$86$	& $-1.835893435$	& $-1.834280747$	& $0.088$ \\ \hline
$88$	& $-1.836277784$	& $-1.834546891$	& $0.094$ \\ \hline
$90$	& $-1.836645303$	& $-1.834974765$	& $0.091$ \\ \hline
$92$	& $-1.836997076$	& $-1.83532337$	& $0.091$ \\ \hline
$94$	& $-1.837334093$	& $-1.835628894$	& $0.093$ \\ \hline
$96$	& $-1.837657262$	& $-1.835970807$	& $0.092$ \\ \hline
$98$	& $-1.837967421$	& $-1.836269121$	& $0.092$ \\ \hline
$100$	& $-1.83826534$	& $-1.836573955$	& $0.092$ \\ \hline
  \end{tabular}
    \end{minipage} 
    \caption{Excess energy for $N=3$.}
  \label{excessEnergyN3Table}
\end{table}
\newpage
\begin{table}[!htb]    
    \begin{minipage}{.5\linewidth}
        \centering
        \begin{tabular}{ | c | c | c | c | }
    \hline
    \hline
    $\Gamma$ & $U_{exc}/q^2 \mbox{ Exact}$ & $U_{exc}/q^2 \mbox{ MC}$ & $\mbox{Error \%}$ \\ \hline
    \hline
    \hline        
$2$	& $-1.841768858$	& $-1.830744844$	& $0.6$ \\ \hline
$4$	& $-2.088755385$	& $-2.075179615$	& $0.65$ \\ \hline
$6$	& $-2.201642808$	& $-2.185315803$	& $0.74$ \\ \hline
$8$	& $-2.267893605$	& $-2.250553847$	& $0.76$ \\ \hline
$10$	& $-2.312255922$	& $-2.292397322$	& $0.86$ \\ \hline
$12$	& $-2.344429254$	& $-2.329785284$	& $0.62$ \\ \hline
$14$	& $-2.369004466$	& $-2.354041135$	& $0.63$ \\ \hline
$16$	& $-2.388451613$	& $-2.376142883$	& $0.52$ \\ \hline
$18$	& $-2.404232879$	& $-2.392897585$	& $0.47$ \\ \hline
$20$	& $-2.417280952$	& $-2.40825754$	& $0.37$ \\ \hline
$22$	& $-2.428227093$	& $-2.419505848$	& $0.36$ \\ \hline
$24$	& $-2.437518974$	& $-2.428899564$	& $0.35$ \\ \hline
$26$	& $-2.445485809$	& $-2.43674796$	& $0.36$ \\ \hline
$28$	& $-2.452376533$	& $-2.44640328$	& $0.24$ \\ \hline
$30$	& $-2.458383433$	& $-2.452283483$	& $0.25$ \\ \hline
$32$	& $-2.463657494$	& $-2.457575282$	& $0.25$ \\ \hline
$34$	& $-2.468318814$	& $-2.462378889$	& $0.24$ \\ \hline
$36$	& $-2.472463921$	& $-2.468327237$	& $0.17$ \\ \hline
$38$	& $-2.476171087$	& $-2.47167007$	& $0.18$ \\ \hline
$40$	& $-2.479504265$	& $-2.47513089$	& $0.18$ \\ \hline
$42$	& $-2.482516075$	& $-2.478046967$	& $0.18$ \\ \hline
$44$	& $-2.485250086$	& $-2.48082382$	& $0.18$ \\ \hline
$46$	& $-2.487742597$	& $-2.48334811$	& $0.18$ \\ \hline
$48$	& $-2.490024024$	& $-2.485557484$	& $0.18$ \\ \hline
$50$	& $-2.492119999$	& $-2.487629366$	& $0.18$ \\ \hline
   \end{tabular}
    \end{minipage}%
    \begin{minipage}{.5\linewidth}
      \centering
        \begin{tabular}{ | c | c | c | c | }
    \hline
    \hline
    $\Gamma$ & $U_{exc}/q^2 \mbox{ Exact}$ & $U_{exc}/q^2 \mbox{ MC}$ & $\mbox{Error \%}$ \\ \hline
    \hline
    \hline    
$52$	& $-2.494052242$	& $-2.489628377$	& $0.18$ \\ \hline
$54$	& $-2.495839254$	& $-2.491391338$	& $0.18$ \\ \hline
$56$	& $-2.497496876$	& $-2.493164385$	& $0.17$ \\ \hline
$58$	& $-2.499038741$	& $-2.494711594$	& $0.17$ \\ \hline
$60$	& $-2.500476639$	& $-2.497806028$	& $0.11$ \\ \hline
$62$	& $-2.50182081$	& $-2.499005918$	& $0.11$ \\ \hline
$64$	& $-2.503080194$	& $-2.500265527$	& $0.11$ \\ \hline
$66$	& $-2.504262621$	& $-2.501483007$	& $0.11$ \\ \hline
$68$	& $-2.505374986$	& $-2.502547918$	& $0.11$ \\ \hline
$70$	& $-2.506423378$	& $-2.503611018$	& $0.11$ \\ \hline
$72$	& $-2.507413196$	& $-2.504507536$	& $0.12$ \\ \hline
$74$	& $-2.508349246$	& $-2.505503412$	& $0.11$ \\ \hline
$76$	& $-2.509235819$	& $-2.506437576$	& $0.11$ \\ \hline
$78$	& $-2.510076755$	& $-2.507196718$	& $0.11$ \\ \hline
$80$	& $-2.51087551$	& $-2.508110117$	& $0.11$ \\ \hline
$82$	& $-2.511635193$	& $-2.508849464$	& $0.11$ \\ \hline
$84$	& $-2.512358614$	& $-2.509588715$	& $0.11$ \\ \hline
$86$	& $-2.513048321$	& $-2.51033724$	& $0.11$ \\ \hline
$88$	& $-2.513706624$	& $-2.510888092$	& $0.11$ \\ \hline
$90$	& $-2.514335628$	& $-2.511602702$	& $0.11$ \\ \hline
$92$	& $-2.514937252$	& $-2.512264073$	& $0.11$ \\ \hline
$94$	& $-2.515513252$	& $-2.512775715$	& $0.11$ \\ \hline
$96$	& $-2.516065234$	& $-2.5133417$	& $0.11$ \\ \hline
$98$	& $-2.516594673$	& $-2.513879685$	& $0.11$ \\ \hline
$100$	& $-2.517102927$	& $-2.51439714$	& $0.11$ \\ \hline
  \end{tabular}
    \end{minipage} 
    \caption{Excess energy for $N=4$.}
  \label{excessEnergyN4Table}
\end{table}
\newpage
\begin{table}[!htb]    
    \begin{minipage}{.5\linewidth}
        \centering
        \begin{tabular}{ | c | c | c | c | }
    \hline
    \hline
    $\Gamma$ & $U_{exc}/q^2 \mbox{ Exact}$ & $U_{exc}/q^2 \mbox{ MC}$ & $\mbox{Error \%}$ \\ \hline
    \hline
    \hline        
$2$&  $-2.273281633$&  $-2.254544383$&  $0.82$  \\ \hline
$4$&  $-2.603059468$&  $-2.57944442$&  $0.91$  \\ \hline
$6$&  $-2.751475064$&  $-2.726853107$&  $0.89$  \\ \hline
$8$&  $-2.836983626$&  $-2.812631384$&  $0.86$  \\ \hline
$10$&  $-2.892752307$&  $-2.86560089$&  $0.94$  \\ \hline
$12$&  $-2.931974479$&  $-2.912520719$&  $0.66$  \\ \hline
$14$&  $-2.961022292$&  $-2.942249886$&  $0.63$  \\ \hline
$16$&  $-2.983373015$&  $-2.968315858$&  $0.5$  \\ \hline
$18$&  $-3.001088911$&  $-2.98650841$&  $0.49$  \\ \hline
$20$&  $-3.015469587$&  $-3.003817336$&  $0.39$  \\ \hline
$22$&  $-3.027373306$&  $-3.016134828$&  $0.37$  \\ \hline
$24$&  $-3.037388678$&  $-3.025977547$&  $0.38$  \\ \hline
$26$&  $-3.045932265$&  $-3.03512191$&  $0.35$  \\ \hline
$28$&  $-3.053306847$&  $-3.0580041$&  $0.15$  \\ \hline
$30$&  $-3.059737633$&  $-3.051782356$&  $0.26$  \\ \hline
$32$&  $-3.065395529$&  $-3.057964147$&  $0.24$  \\ \hline
$34$&  $-3.07041254$&  $-3.062483839$&  $0.26$  \\ \hline
$36$&  $-3.074892226$&  $-3.068935758$&  $0.19$  \\ \hline
$38$&  $-3.078916966$&  $-3.073237404$&  $0.18$  \\ \hline
$40$&  $-3.082553109$&  $-3.076951143$&  $0.18$  \\ \hline
   \end{tabular}
    \end{minipage}%
    \begin{minipage}{.5\linewidth}
      \centering
        \begin{tabular}{ | c | c | c | c | }
    \hline
    \hline
    $\Gamma$ & $U_{exc}/q^2 \mbox{ Exact}$ & $U_{exc}/q^2 \mbox{ MC}$ & $\mbox{Error \%}$ \\ \hline
    \hline
    \hline    
$42$&  $-3.08585469$&  $-3.080149174$&  $0.18$  \\ \hline
$44$&  $-3.088866161$&  $-3.083299187$&  $0.18$  \\ \hline
$46$&  $-3.091624427$&  $-3.104185138$&  $0.41$  \\ \hline
$48$&  $-3.094160381$&  $-3.088421905$&  $0.19$  \\ \hline
$50$&  $-3.096500081$&  $-3.090806996$&  $0.18$  \\ \hline
$52$&  $-3.098665665$&  $-3.093139164$&  $0.18$  \\ \hline
$54$&  $-3.100676054$&  $-3.095168147$&  $0.18$  \\ \hline
$56$&  $-3.102547523$&  $-3.097061871$&  $0.18$  \\ \hline
$58$&  $-3.104294145$&  $-3.09900577$&  $0.17$  \\ \hline
$60$&  $-3.105928151$&  $-3.102281958$&  $0.12$  \\ \hline
$62$&  $-3.107460225$&  $-3.103852903$&  $0.12$  \\ \hline
$64$&  $-3.108899737$&  $-3.1052599$&  $0.12$  \\ \hline
$66$&  $-3.110254943$&  $-3.106666443$&  $0.12$  \\ \hline
$68$&  $-3.111533142$&  $-3.107837135$&  $0.12$  \\ \hline
$70$&  $-3.112740814$&  $-3.109142488$&  $0.12$  \\ \hline
$72$&  $-3.113883731$&  $-3.110264541$&  $0.12$  \\ \hline
$74$&  $-3.114967051$&  $-3.111243733$&  $0.12$  \\ \hline
$76$&  $-3.115995396$&  $-3.112546659$&  $0.11$  \\ \hline
$78$&  $-3.116972923$&  $-3.113479064$&  $0.11$  \\ \hline
$80$&  $-3.117903377$&  $-3.114304072$&  $0.12$  \\ \hline 
  \end{tabular}
    \end{minipage} 
    \caption{Excess energy for $N=5$.}
  \label{excessEnergyN5Table}
\end{table}

\begin{table}[!htb]    
    \begin{minipage}{.5\linewidth}
        \centering
        \begin{tabular}{ | c | c | c | c | }
    \hline
    \hline
    $\Gamma$ & $U_{exc}/q^2 \mbox{ Exact}$ & $U_{exc}/q^2 \mbox{ MC}$ & $\mbox{Error \%}$ \\ \hline
    \hline
    \hline        
$2$&  $-2.704455625$&  $-2.676202387$&  $1.$  \\ \hline
$4$&  $-3.117236507$&  $-3.084759772$&  $1.$  \\ \hline
$6$&  $-3.301910232$&  $-3.270457533$&  $0.95$  \\ \hline
$8$&  $-3.40880686$&  $-3.376551757$&  $0.95$  \\ \hline
$10$&  $-3.479519106$&  $-3.446610638$&  $0.95$  \\ \hline
$12$&  $-3.530347907$&  $-3.505801688$&  $0.7$  \\ \hline
$14$&  $-3.569010404$&  $-3.544929773$&  $0.67$  \\ \hline
$16$&  $-3.599635921$&  $-3.580374885$&  $0.54$  \\ \hline
$18$&  $-3.624636756$&  $-3.605402871$&  $0.53$  \\ \hline
$20$&  $-3.645519798$&  $-3.63025116$&  $0.42$  \\ \hline
$22$&  $-3.663277844$&  $-3.64829552$&  $0.41$  \\ \hline
$24$&  $-3.67859409$&  $-3.664193612$&  $0.39$  \\ \hline
$26$&  $-3.691956074$&  $-3.677636723$&  $0.39$  \\ \hline
$28$&  $-3.703722589$&  $-3.692785401$&  $0.3$  \\ \hline
$30$&  $-3.714164751$&  $-3.703978618$&  $0.27$  \\ \hline
  \end{tabular}
    \end{minipage}%
    \begin{minipage}{.5\linewidth}
      \centering
        \begin{tabular}{ | c | c | c | c | }
    \hline
    \hline
    $\Gamma$ & $U_{exc}/q^2 \mbox{ Exact}$ & $U_{exc}/q^2 \mbox{ MC}$ & $\mbox{Error \%}$ \\ \hline
    \hline
    \hline    
$32$&  $-3.723492177$&  $-3.713024923$&  $0.28$  \\ \hline
$34$&  $-3.731870209$&  $-3.721623116$&  $0.27$  \\ \hline
$36$&  $-3.739431585$&  $-3.731761699$&  $0.21$  \\ \hline
$38$&  $-3.746284526$&  $-3.738276826$&  $0.21$  \\ \hline
$40$&  $-3.752518482$&  $-3.744793309$&  $0.21$  \\ \hline
$42$&  $-3.758208275$&  $-3.750743175$&  $0.2$  \\ \hline
$44$&  $-3.76341715$&  $-3.755970983$&  $0.2$  \\ \hline
$46$&  $-3.768199061$&  $-3.760613314$&  $0.2$  \\ \hline
$48$&  $-3.772600394$&  $-3.765284346$&  $0.19$  \\ \hline
$50$&  $-3.776661309$&  $-3.769001687$&  $0.2$  \\ \hline
$52$&  $-3.780416776$&  $-3.773516744$&  $0.18$  \\ \hline
$54$&  $-3.7838974$&  $-3.776585328$&  $0.19$  \\ \hline
$56$&  $-3.787130083$&  $-3.780115258$&  $0.19$  \\ \hline
$58$&  $-3.790138561$&  $-3.783557231$&  $0.17$  \\ \hline
$60$&  $-3.792943845$&  $-3.788307793$&  $0.12$  \\ \hline
  \end{tabular}
    \end{minipage} 
    \caption{Excess energy for $N=6$.}
  \label{excessEnergyN6Table}
\end{table}
\newpage

\begin{table}[!htb]    
    \begin{minipage}{.5\linewidth}
        \centering
        \begin{tabular}{ | c | c | c | c | }
    \hline
    \hline
    $\Gamma$ & $U_{exc}/q^2 \mbox{ Exact}$ & $U_{exc}/q^2 \mbox{ MC}$ & $\mbox{Error \%}$ \\ \hline
    \hline
    \hline        
$2$&  $-3.135434539$&  $-3.099996871$&  $1.1$  \\ \hline
$4$&  $-3.631312057$&  $-3.59179127$&  $1.1$  \\ \hline
$6$&  $-3.852046648$&  $-3.812545183$&  $1.$  \\ \hline
$8$&  $-3.97942912$&  $-3.942828269$&  $0.92$  \\ \hline
$10$&  $-4.063212623$&  $-4.026786936$&  $0.9$  \\ \hline
$12$&  $-4.122849606$&  $-4.094647925$&  $0.68$  \\ \hline
$14$&  $-4.167592063$&  $-4.141706971$&  $0.62$  \\ \hline
$16$&  $-4.202438377$&  $-4.180506329$&  $0.52$  \\ \hline
$18$&  $-4.230344658$&  $-4.209768458$&  $0.49$  \\ \hline
$20$&  $-4.253181903$&  $-4.236646822$&  $0.39$  \\ \hline
$22$&  $-4.272198258$&  $-4.25610523$&  $0.38$  \\ \hline
$24$&  $-4.28826176$&  $-4.273232198$&  $0.35$  \\ \hline
$26$&  $-4.301996594$&  $-4.287218267$&  $0.34$  \\ \hline
  \end{tabular}
    \end{minipage}%
    \begin{minipage}{.5\linewidth}
      \centering
        \begin{tabular}{ | c | c | c | c | }
    \hline
    \hline
    $\Gamma$ & $U_{exc}/q^2 \mbox{ Exact}$ & $U_{exc}/q^2 \mbox{ MC}$ & $\mbox{Error \%}$ \\ \hline
    \hline
    \hline    
$28$&  $-4.313863936$&  $-4.302615303$&  $0.26$  \\ \hline
$30$&  $-4.324212197$&  $-4.313186123$&  $0.25$  \\ \hline
$32$&  $-4.333309524$&  $-4.32269442$&  $0.24$  \\ \hline
$34$&  $-4.341365562$&  $-4.330944347$&  $0.24$  \\ \hline
$36$&  $-4.348546447$&  $-4.339962497$&  $0.2$  \\ \hline
$38$&  $-4.354985403$&  $-4.346585279$&  $0.19$  \\ \hline
$40$&  $-4.360790395$&  $-4.352554547$&  $0.19$  \\ \hline
$42$&  $-4.366049748$&  $-4.358110539$&  $0.18$  \\ \hline
$44$&  $-4.370836348$&  $-4.362931534$&  $0.18$  \\ \hline
$46$&  $-4.375210821$&  $-4.367216025$&  $0.18$  \\ \hline
$48$&  $-4.379223966$&  $-4.371833475$&  $0.17$  \\ \hline
$50$&  $-4.382918643$&  $-4.375658727$&  $0.17$  \\ \hline
$ $&  $ 		$&  $		$&  $	$  \\ \hline
  \end{tabular}
    \end{minipage} 
    \caption{Excess energy for $N=7$.}
  \label{excessEnergyN7Table}
\end{table}

\begin{table}[!htb]    
    \begin{minipage}{.5\linewidth}
        \centering
        \begin{tabular}{ | c | c | c | c | }
    \hline
    \hline
    $\Gamma$ & $U_{exc}/q^2 \mbox{ Exact}$ & $U_{exc}/q^2 \mbox{ MC}$ & $\mbox{Error \%}$ \\ \hline
    \hline
    \hline        
$2$&  $-3.566290974$&  $-3.523322613$&  $1.2$  \\ \hline
$4$&  $-4.145324917$&  $-4.097784$&  $1.1$  \\ \hline
$6$&  $-4.402127425$&  $-4.357011378$&  $1.$  \\ \hline
$8$&  $-4.550132893$&  $-4.509247701$&  $0.9$  \\ \hline
$10$&  $-4.64743959$&  $-4.607689971$&  $0.86$  \\ \hline
$12$&  $-4.716754014$&  $-4.68628409$&  $0.65$  \\ \hline
$14$&  $-4.768881725$&  $-4.740471927$&  $0.6$  \\ \hline
$16$&  $-4.809655179$&  $-4.786774034$&  $0.48$  \\ \hline
  \end{tabular}
    \end{minipage}%
    \begin{minipage}{.5\linewidth}
      \centering
        \begin{tabular}{ | c | c | c | c | }
    \hline
    \hline
    $\Gamma$ & $U_{exc}/q^2 \mbox{ Exact}$ & $U_{exc}/q^2 \mbox{ MC}$ & $\mbox{Error \%}$ \\ \hline
    \hline
    \hline    
$18$&  $-4.842510641$&  $-4.821024596$&  $0.44$  \\ \hline
$20$&  $-4.869607597$&  $-4.851542428$&  $0.37$  \\ \hline
$22$&  $-4.892372715$&  $-4.875037791$&  $0.35$  \\ \hline
$24$&  $-4.911787394$&  $-4.894623458$&  $0.35$  \\ \hline
$26$&  $-4.928549963$&  $-4.912176512$&  $0.33$  \\ \hline
$28$&  $-4.943171967$&  $-4.93066387$&  $0.25$  \\ \hline
$30$&  $-4.956037721$&  $-4.943661916$&  $0.25$  \\ \hline
$ $&  $ 		$&  $		$&  $	$  \\ \hline
  \end{tabular}
    \end{minipage} 
    \caption{Excess energy for $N=8$.}
  \label{excessEnergyN8Table}
\end{table}

\begin{table}[!htb]    
    \begin{minipage}{.5\linewidth}
        \centering
        \begin{tabular}{ | c | c | c | c | }
    \hline
    \hline
    $\Gamma$ & $U_{exc}/q^2 \mbox{ Exact}$ & $U_{exc}/q^2 \mbox{ MC}$ & $\mbox{Error \%}$ \\ \hline
    \hline
    \hline        
$2$&  $-3.997065516$&  $-3.944975802$&  $1.3$  \\ \hline
$4$&  $-4.659297158$&  $-4.604135518$&  $1.2$  \\ \hline
$6$&  $-4.952187108$&  $-4.900700727$&  $1.$  \\ \hline
$8$&  $-5.120892865$&  $-5.075166478$&  $0.89$  \\ \hline
  \end{tabular}
    \end{minipage}%
    \begin{minipage}{.5\linewidth}
      \centering
        \begin{tabular}{ | c | c | c | c | }
    \hline
    \hline
    $\Gamma$ & $U_{exc}/q^2 \mbox{ Exact}$ & $U_{exc}/q^2 \mbox{ MC}$ & $\mbox{Error \%}$ \\ \hline
    \hline
    \hline    
$10$&  $-5.231864477$&  $-5.190094538$&  $0.8$  \\ \hline
$12$&  $-5.311018678$&  $-5.278356343$&  $0.61$  \\ \hline
$14$&  $-5.370657753$&  $-5.340463742$&  $0.56$  \\ \hline
$16$&  $-5.417405306$&  $-5.392460678$&  $0.46$  \\ \hline
  \end{tabular}
    \end{minipage} 
    \caption{Excess energy for $N=9$.}
  \label{excessEnergyN9Table}
\end{table}

\newpage
\section{Excess energy as a function of $N$}
\label{app:UexcNfit}

In this section, we report the excess energy $U_{exc}$ obtained from
the exact expression, Eq.~(\ref{exactUexcAnyGammaEq}), when $N$
increases, for three fixed values of $\Gamma=4, 6, 8$. As before, we
have set $\rho_b=1$ and $L=1$. A four parameter fit to an ansatz of the form
\begin{equation}
   U_{exc}=q^2(A N + B + C/N + D/N^2) 
\end{equation}
is proposed. As explained in Sec.~\ref{subsec:energyNlarge}, this is
the expected finite-size expansion for the excess energy. The fit is
done with four consecutive values of $N$, and the convergence of the
parameters $A$, $B$, $C$ and $D$ is observed as $N$ increases. This
allows us to obtain the bulk value of the excess internal energy and the
finite size corrections.

\newpage


\begin{table}[!htb]
  \begin{center}
\begin{tabular}{|c|c|c|c|c|c|c|}
\hline
{$\Gamma$} & {$N$} & {$U_{exc}/q^2 \mbox{ Exact}$} &{$A$} & {$B$} & {$C$} & {$D$} \\ \hline
\multicolumn{1}{|l|}{} & \multicolumn{1}{l|}{} & \multicolumn{1}{l|}{} & \multicolumn{1}{l|}{} & \multicolumn{1}{l|}{} & \multicolumn{1}{l|}{} & \multicolumn{1}{l|}{} \\ \hline
4 & 2 & -1.0591247 & \multicolumn{1}{l|}{} & \multicolumn{1}{l|}{} & \multicolumn{1}{l|}{} & \multicolumn{1}{l|}{} \\ \hline
4 & 3 & -1.5738739 & \multicolumn{1}{l|}{} & \multicolumn{1}{l|}{} & \multicolumn{1}{l|}{} & \multicolumn{1}{l|}{} \\ \hline
4 & 4 & -2.0887554 & \multicolumn{1}{l|}{} & \multicolumn{1}{l|}{} & \multicolumn{1}{l|}{} & \multicolumn{1}{l|}{} \\ \hline
4 & 5 & -2.6030595 & -0.5123875 & -0.0519396 & 0.0666943 & -0.0630289 \\ \hline
4 & 6 & -3.1172365 & -0.5144079 & -0.0276949 & -0.0282639 & 0.0581944 \\ \hline
4 & 7 & -3.6313121 & -0.5136870 & -0.0385086 & 0.0250833 & -0.0283146 \\ \hline
4 & 8 & -4.1453249 & -0.5138297 & -0.0359394 & 0.0098110 & 0.0016590 \\ \hline
4 & 9 & -4.6592972 & -0.5138406 & -0.0357102 & 0.0082173 & 0.0053268 \\ \hline
4 & 10 & -5.1732408 & -0.5138274 & -0.0360269 & 0.0107380 & -0.0013247 \\ \hline
4 & 11 & -5.6871636 & -0.5138275 & -0.0360265 & 0.0107347 & -0.0013147 \\ \hline
4 & 12 & -6.2010707 & -0.5138292 & -0.0359730 & 0.0102012 & 0.0004514 \\ \hline
4 & 13 & -6.7149658 & -0.5138291 & -0.0359789 & 0.0102654 & 0.0002174 \\ \hline
4 & 14 & -7.2288514 & -0.5138290 & -0.0359827 & 0.0103120 & 0.0000320 \\ \hline
\multicolumn{1}{|l|}{} & \multicolumn{1}{l|}{} & \multicolumn{1}{l|}{} & \multicolumn{1}{l|}{} & \multicolumn{1}{l|}{} & \multicolumn{1}{l|}{} & \multicolumn{1}{l|}{} \\ \hline
6 & 2 & -1.1007914 & \multicolumn{1}{l|}{} & \multicolumn{1}{l|}{} & \multicolumn{1}{l|}{} & \multicolumn{1}{l|}{} \\ \hline
6 & 3 & -1.6494692 & \multicolumn{1}{l|}{} & \multicolumn{1}{l|}{} & \multicolumn{1}{l|}{} & \multicolumn{1}{l|}{} \\ \hline
6 & 4 & -2.2016428 & \multicolumn{1}{l|}{} & \multicolumn{1}{l|}{} & \multicolumn{1}{l|}{} & \multicolumn{1}{l|}{} \\ \hline
6 & 5 & -2.7514751 & -0.5400872 & -0.1087176 & 0.3631889 & -0.3739746 \\ \hline
6 & 6 & -3.3019102 & -0.5569618 & 0.0937774 & -0.4299168 & 0.6385007 \\ \hline
6 & 7 & -3.8520466 & -0.5463876 & -0.0648361 & 0.3525767 & -0.6304077 \\ \hline
6 & 8 & -4.4021274 & -0.5507877 & 0.0143664 & -0.1182382 & 0.2936214 \\ \hline
6 & 9 & -4.9521871 & -0.5501298 & 0.0005507 & -0.0221860 & 0.0725700 \\ \hline
6 & 10 & -5.5022159 & -0.5497171 & -0.0093559 & 0.0566539 & -0.1354685 \\ \hline
6 & 11 & -6.0522211 & -0.5498824 & -0.0048913 & 0.0166380 & -0.0164129 \\ \hline
6 & 12 & -6.6022125 & -0.5499920 & -0.0016029 & -0.0161368 & 0.0921056 \\ \hline
6 & 13 & -7.1521927 & -0.5499078 & -0.0043821 & 0.0143501 & -0.0190619 \\ \hline
6 & 14 & -7.7021636 & -0.5499020 & -0.0045916 & 0.0168585 & -0.0290490 \\ \hline
\multicolumn{1}{|l|}{} & \multicolumn{1}{l|}{} & \multicolumn{1}{l|}{} & \multicolumn{1}{l|}{} & \multicolumn{1}{l|}{} & \multicolumn{1}{l|}{} & \multicolumn{1}{l|}{} \\ \hline
8 & 2 & -1.1257914 & \multicolumn{1}{l|}{} & \multicolumn{1}{l|}{} & \multicolumn{1}{l|}{} & \multicolumn{1}{l|}{} \\ \hline
8 & 3 & -1.6928841 & \multicolumn{1}{l|}{} & \multicolumn{1}{l|}{} & \multicolumn{1}{l|}{} & \multicolumn{1}{l|}{} \\ \hline
8 & 4 & -2.2678936 & \multicolumn{1}{l|}{} & \multicolumn{1}{l|}{} & \multicolumn{1}{l|}{} & \multicolumn{1}{l|}{} \\ \hline
8 & 5 & -2.8369836 & -0.5450671 & -0.2531712 & 0.8893389 & -0.9086220 \\ \hline
8 & 6 & -3.4088069 & -0.5943690 & 0.3384527 & -1.4278551 & 2.0494980 \\ \hline
8 & 7 & -3.9794291 & -0.5547255 & -0.2562003 & 1.5057664 & -2.7077260 \\ \hline
8 & 8 & -4.5501329 & -0.5764972 & 0.1356898 & -0.8238020 & 1.8643240 \\ \hline
8 & 9 & -5.1208929 & -0.5709729 & 0.0196798 & -0.0172566 & 0.0081648 \\ \hline
8 & 10 & -5.6915851 & -0.5691710 & -0.0235650 & 0.3268999 & -0.8999760 \\ \hline
8 & 11 & -6.2622272 & -0.5704051 & 0.0097535 & 0.0282679 & -0.0114840 \\ \hline
\end{tabular}
  \end{center}
\caption{Excess internal energy of the OCP in the sphere and its fit to  $U_{exc}=q^2(A N + B + C/N + D/N^2)$.}
\label{tab:UintG4_6_8}
\end{table}


\newpage

\end{document}